\documentclass[%
longbibliography,
 amsmath,amssymb,
onecolumn,
]{revtex4-2}

\usepackage{graphicx}
\usepackage{dcolumn}
\usepackage{hyperref}
\usepackage{bm}

\begin{document}

\preprint{APS/123-QED}

\title{Dynamics-based machine learning of transitions in Couette flow}

\author{B\'alint Kasz\'as}
\author{Mattia Cenedese}%
 \author{George Haller}
 \email[]{georgehaller@ethz.ch}
\affiliation{Institute for Mechanical Systems, ETH Z\"urich, Leonhardstrasse 21, 8092 Z\"urich, Switzerland}%

\date{August 25, 2022}
\begin{abstract}
We derive low-dimensional, data-driven  models for transitions among exact coherent states (ECSs) in one of the most studied canonical shear flows, the plane Couette flow. These one- or two-dimensional nonlinear models represent the leading-order reduced dynamics on attracting spectral submanifolds (SSMs), which we construct using the recently developed \texttt{SSMLearn} algorithm from a small number of simulated transitions. We find that the energy input and dissipation rates provide  efficient parametrizations for the most important SSMs. By restricting the dynamics to these SSMs, we obtain reduced-order models that also reliably predict nearby, off-SSM  transitions that were not used in their training. 
\end{abstract}
\maketitle
\section{Introduction}
A detailed analysis of complex nonlinear dynamical systems, such as those arising in fluid mechanics, generally requires reduced-order models. The two main reduction techniques available for this purpose are the proper orthogonal decomposition (POD)  \cite{holmesTurbulenceCoherentStructures1996} and the dynamic mode decomposition (DMD) \cite{schmidDMDRev2022}. The POD approach is  equation-driven, projecting the governing equations onto an empirically selected set of most energetic modes. In contrast, the DMD approach is data-driven, fitting a linear dynamical system to the dynamics of a set of observables. Neither of these approaches is, therefore, designed for a purely data-driven modeling of essentially nonlinear (or non-linearizable) behavior \cite{pageKoopmanModeExpansions2019}.  

{ A hallmark of non-linearizability on a domain of the phase space of a dynamical system is the coexistence of isolated stationary states (called invariant solutions in the fluid dynamics literature \cite{kawahara_significance_2012}) and transitions among them, which are as ubiquitous in  laminar Couette flows as in turbulent pipe flows.} Several emerging machine learning techniques could formally be applied to pattern-match transitions in such flows. These techniques, however,  are not yet mature enough to produce physically interpretable models of reasonable complexity that can reliably  predict the dynamics for initial conditions not used in their training \cite{bruntonRev2020}.

{ A recent approach reduces non-linearizable dynamics to spectral submanifolds (SSMs) \cite{hallerNonlinearNormalModes2016a, kogelbauerRigorousModelReduction2018}, which are the smoothest nonlinear continuations of invariant (spectral) subspaces of the linearized system a stationary state, such as a fixed point, a periodic orbits, or a quasiperiodic torus \cite{hallerNonlinearNormalModes2016a}. Building on more abstract prior work by \cite{cabreParameterizationMethodInvariant2003a,haro2006}, SSM theory establishes the existence and uniqueness of spectral submanifolds if the eigenvalues corresponding to the spectral subspace are not in resonance with the rest of the spectrum. See Supplemental Material \cite{supplementalmaterial} at [URL will be inserted by publisher] for more detail. In the vicinity of the smoothest (primary) SSM, other, less smooth (secondary) SSMs also exist. Coexisting stationary states might also be contained in such secondary SSMs \cite{cenedeseDatadrivenModelingPrediction2022} in a vicinity of the smoothest SSM, as indeed turns out to be the case here.} Restricting the governing equations to such low-dimensional attracting SSMs, therefore, provides a fast and mathematically exact model reduction procedure, as has been demonstrated for finite-element models of beams, shells and wings, with degrees of freedom ranging up to hundreds of thousands \cite{jainHowComputeInvariant2022}. {Due to the invariance of the SSM used in the model order reduction, the accuracy of the computed models depends only on the accuracy of the numerical approximation for the SSMs. This can be gradually enhanced by increasing the polynomial order of expansion for the SSM without increasing the dimension of the reduced model. SSM-based reduction has yielded highly accurate and predictive reduced models of very high dimensional finite-element equations  \cite{jainHowComputeInvariant2022,li2022}.}

Very recently, SSM-based model reduction has been extended to a fully data-driven setting \cite{cenedeseDatadrivenModelingPrediction2022}. Implemented in the open-source \texttt{SSMLearn} package, the algorithm identifies the most influential nonresonant spectral subspaces near a stationary state from a linear modal analysis of a few training trajectories. It then proceeds to  construct the corresponding nonresonant SSMs and {the extended} normal form of their reduced dynamics from the training data. The resulting SSM-based models have proven to be accurate in predicting behavior in several problems from solid and fluid mechanics, even under the addition of external forcing absent in their training data \cite{cenedeseDatadrivenModelingPrediction2022,cenedeseDatadrivenModelingPrediction2022b,axas2022}.

It has been unclear, however, whether data-driven SSM-reduction can also describe  transitions among coexisting stationary states accurately. 
Such phenomena are omnipresent in a number of outstanding problems of applied science, such as transition to turbulence \cite{kawahara_significance_2012} and tipping points \cite{lenton_tipping2008} in climate. To examine the applicability of the SSM-based approach to these grand challenges, it is feasible start with similar but simpler canonical problems that are nevertheless of significant interest in their own right. In this paper, we will carry out such an exploratory study for one of the most studied classic shear flows, the plane Couette flow. 

The anchor points of SSM theory, stationary states, have been broadly studied in the Couette flow literature as exact coherent states (ECSs) \cite{waleffeExactCoherentStructures2001,gibsonVisualizingGeometryState2008}. These can be  equilibrium points, traveling waves, periodic orbits, quasi-periodic trajectories or even chaotic attractors. Transitions between ECSs are generally understood to happen along  their stable and unstable manifolds \cite{fujimuraCenter1997,gibsonVisualizingGeometryState2008,cariniCentremanifoldReductionBifurcating2015,budanurUnstableManifoldsRelative2017, budanurHeteroclinicPathSpatially2017a, faranoComputingHeteroclinicOrbits2019}. This is even supported by experimental observations \cite{suriForecastingFluidFlows2017}. Yet,  with the exception of the two-dimensional unstable manifold of the base state in the flow past a cylinder \cite{loiseauPODmanifold2020},  the role of invariant manifolds emanating from ECSs has not yet been systematically explored in model  reduction.

An added motivation 
is recent work on the applicability of linear data-driven modeling tools  to transitions between ECSs in plane Couette flow \cite{pageKoopmanModeExpansions2019}. The conclusion of that study is that  while DMD can feature-match individual transition trajectories over certain time intervals, no underlying convergent Koopman mode decomposition \cite{rowleySpectralAnalysisNonlinear2009} justifying such a formal DMD analysis  exists beyond subsets of the domains of attraction or repulsion of the ECSs. In more practical terms, while linear model reduction methods fitted to non-linearizable data sets always return a closest-fitting linear model, the predictive power of such a model is limited to an a priori unknown subset of the phase space that contains a single stable or unstable ECS. Capturing multiple ECSs and the transitions among them in a single reduced-order model for Couette flows has been, therefore, an outstanding challenge which we wish to address here using data-driven, SSM-based nonlinear model reduction. Specifically, we seek to compute the most influential SSMs and their reduced dynamics in the phase space 
via the dynamics-based machine learning approach of  \cite{cenedeseDatadrivenModelingPrediction2022}. We rely solely on simulation data, which renders our approach non-intrusive. We parametrize the SSMs with physically interpretable observables used in earlier studies of this flow to obtain very low (one or two) dimensional dynamical systems for these observables along the SSMs. We also assess the predictive power of these reduced models for nearby trajectories starting off the SSMs. {In this Letter, we seek the parametrization of SSMs and their reduced dynamics using classical machine-learning techniques, such as polynomial regression. Based on the mathematical existence results on SSMs, one can also seek SSM-based models with more advanced tools from machine learning \cite{hastie2009elements}.}  We refer to the Supplemental Material \cite{supplementalmaterial} and the Matlab live scripts under \cite{ssmlearnCouette} for more information on our data sets and computations. 

\section{Setup}
We consider the plane Couette flow configuration \cite{orrStabilityInstabilitySteady1907}: an incompressible fluid flow between two infinite plates moving in opposite directions. The velocity field $\mathbf{u} = [u, v, w](x,y,z,t)$ evolves according to the Navier--Stokes equations {along with the pressure $p$}
\begin{equation}
\label{eq:ns}
    \frac{\partial \mathbf{u}}{\partial t} + \mathbf{u}\cdot \nabla \mathbf{u} = -\nabla p + \frac{1}{\text{Re}} \Delta \mathbf{u}, \quad \nabla \cdot \mathbf{u} = 0,
\end{equation}
{in the physical domain ${\Omega = [0, L_x] \times [-h, h] \times [0, L_z]}$, where the Reynolds number is defined as $\text{Re}=Uh/\nu$, with $\nu$ denoting the kinematic viscosity, $h$ is half the distance between the plates, and $U$ is their velocity. Our equations are nondimensionalized by the plate velocity $U$ and the half-width of the channel $h$. } As in previous studies \cite{pageKoopmanModeExpansions2019,gibsonEquilibriumTravellingwaveSolutions2009}, we impose periodic boundary conditions in the $x$ (spanwise) and $z$ (streamwise) directions and in the wall normal direction we require $u(x,y=\pm 1, z, t) = \pm 1$. 
With the choice $L_x = 5\pi /2$ and $L_z =4\pi/3$, our computational cell is comparable to those used in \cite{nagataThreedimensionalFiniteamplitudeSolutions1990,pageKoopmanModeExpansions2019}.

We solve Eq. (\ref{eq:ns}) using the open source  \texttt{Channelflow} library \cite{channelflow}, with a spectral discretization of $32\times35\times32$ modes, resulting in a phase space of dimension $O(10^5)$. This is a very high but still finite-dimensional phase space to which the mathematical results behind \texttt{SSMLearn} are directly applicable.
We rely on \texttt{Channelflow} only for generating trajectories, which we use for training and testing our SSM-based, data-driven models via the open source toolbox \texttt{SSMLearn} \cite{cenedeseDatadrivenModelingPrediction2022}.

Here we focus on the low Reynolds number regime in which the plane Couette flow is bistable \cite{gibsonEquilibriumTravellingwaveSolutions2009}. {In this geometry, at around $\text{Re}=134.5$, the lower- and upper branch of fixed points first observed by Nagata \cite{waleffeHomotopyExactCoherent2003} appear in a saddle-node bifurcation.} 
{We restrict all calculations to the invariant subspace of the group generated by the shift-reflect and shift-rotate symmetries \cite{gibsonVisualizingGeometryState2008}, the isotropy subgroup of the Nagata-equilibria. This restriction ensures that the stationary states we study become hyperbolic, i.e., their spectrum does not contain the zero eigenvalues arising from the transnational invariance of the governing equations. Hyperbolicity of the stationary states is required for all SSM results to apply and also insures the robustness (structural stability) of our results.  All orbits connecting these hyperbolic stationary states also remain in this invariant subspace. }

{Together with the constant-shear base state, $\mathbf{u}_{\text{base}}=[u = y, v=0, w=0]$, the Nagata equilibria represent the simplest examples of ECSs, with transitions among them along SSMs. Specifically, trajectories close to the unstable fixed point evolve along the unstable manifold that connects into slow SSMs of the two attracting states.} 
At higher Re, non-trivial ECSs (e.g., periodic or quasi-periodic orbits) start appearing via bifurcations \cite{kawaharaPeriodicMotionEmbedded2001}. 

Our aim here is to develop reduced models on SSMs that predict transitions between three coexisting ECSs based on a small number of simulated transitions. To parametrize the SSMs,  we choose physically relevant quantities from previous studies of the same flow \cite{kawaharaPeriodicMotionEmbedded2001,gibsonVisualizingGeometryState2008}, such as  the rate of energy input (I) supplied by the walls and the rate of energy dissipation (D) due to friction, often used to get a projected representation of the Navier--Stokes equations \cite{gibsonVisualizingGeometryState2008, vanveenHomoclinicTangleEdge2011, kawaharaPeriodicMotionEmbedded2001}, motivated by an energy balance of the flow \cite{doeringEnergyDissipationShear1992}. Considering the time dependence of the total energy defined as 
\begin{equation}
    E =\frac{||\mathbf{u}(t)||^2_{L^2}}{2} =  \frac{1}{2L_x L_z} \int_0^{L_x}\int_{-1}^1\int_0^{L_z} \frac{|\mathbf{u}(x,y,z,t)|^2}{2} \  \text{d}z \ \text{d}y \ \text{d}x ,
\end{equation}
one can derive the differential equation governing $E(t)$ by taking the inner product of the Navier-Stokes equations with the velocity field $
\mathbf{u}$ and integrating over the flow domain $\Omega$. Due to incompressibility and periodicity, one obtains \cite{waleffeOverviewTurbulentShear2011}
\begin{equation}\label{eq:energybalance}
    \dot{E} = (I - D)/\text{Re},
\end{equation}
\noindent where the rate of energy input $I$ and the dissipation $D$ are defined as 
\begin{align}
    I & = \frac{1}{2L_x L_z}\int_0^{L_x} \int_0^{L_z} \left( \left( \frac{\partial u}{\partial y}u \right)_{y=1} + \left( \frac{\partial u}{\partial y}u \right)_{y=-1} \right)\text{d}z  \ \text{d}x -1,\label{eq:I}\\
    D &=  \frac{1}{2L_x L_z} \int_0^{L_x}\int_{-1}^1\int_0^{L_z}|\nabla \times \mathbf{u}|^2 \  \text{d}z \ \text{d}y \ \text{d}x -1 ,\label{eq:D}
\end{align}
in order to measure these quantities relative to the constant shear base state. Another energy variable we use is $\Delta E = ||\mathbf{u}(t)-\mathbf{u}_{\text{base}}||^2_{L^2} / 2$. Our definitions ensure that $\Delta E=D=I=0$ at the base state. 
\begin{figure}[t!]
    \centering
    \includegraphics[width = 0.5\linewidth]{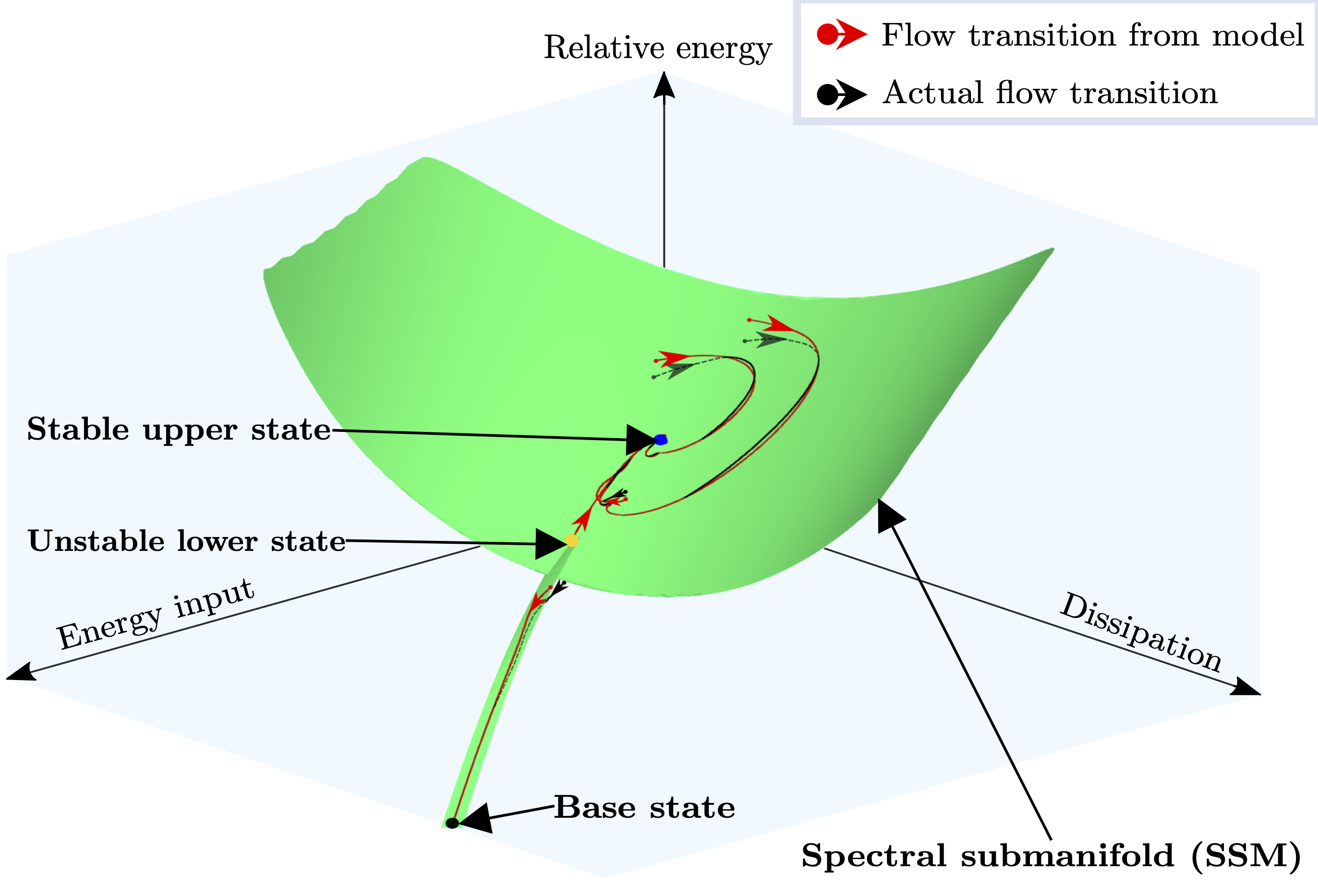}
    \caption{Phase space geometry of plane Couette flow,  projected onto the space of three observables: the relative energy $\Delta E$, the energy input rate $I$ and the energy dissipation rate $D$.   The three coexisting fixed points are the constant-shear base state (black dot), the unstable Nagata lower branch (yellow dot), and the stable Nagata upper branch (blue dot) fixed points. {A two-dimensional SSM approximated by Eq. (7), containing all three fixed points, is shown in green. Due to the large distance between the base state and the lower Nagata equilibrium, we indicate the base state at a higher $\Delta E$ value along the manifold.} Also shown are model-predicted and actual transitions between ECSs.}
    \label{fig:1}
\end{figure}

{Parametrizations depending on $I$ and $D$ suffer from a singularity at the base state $I=D=0$ due to their quadratic-type dependence on the velocity, cf. Eqs. (4-5).
To remedy this problem without losing physical motivation, we choose $J = \sqrt{|I|}$ and  $K = \sqrt{|D|}$ 
as parametrization variables (see the Supplemental Material)}. \texttt{SSMLearn} then reveals a smooth dependence of the reduced dynamics on the SSM graphed over this set of variables. Figure \ref{fig:1} shows an SSM, parametrized by $(J,K)$, plotted in the three-dimensional space $(I, D, \Delta E)$ with its dynamics describing transitions between ECSs.

\section{Results}
\begin{figure*}[t!]
    \includegraphics[width = \textwidth]{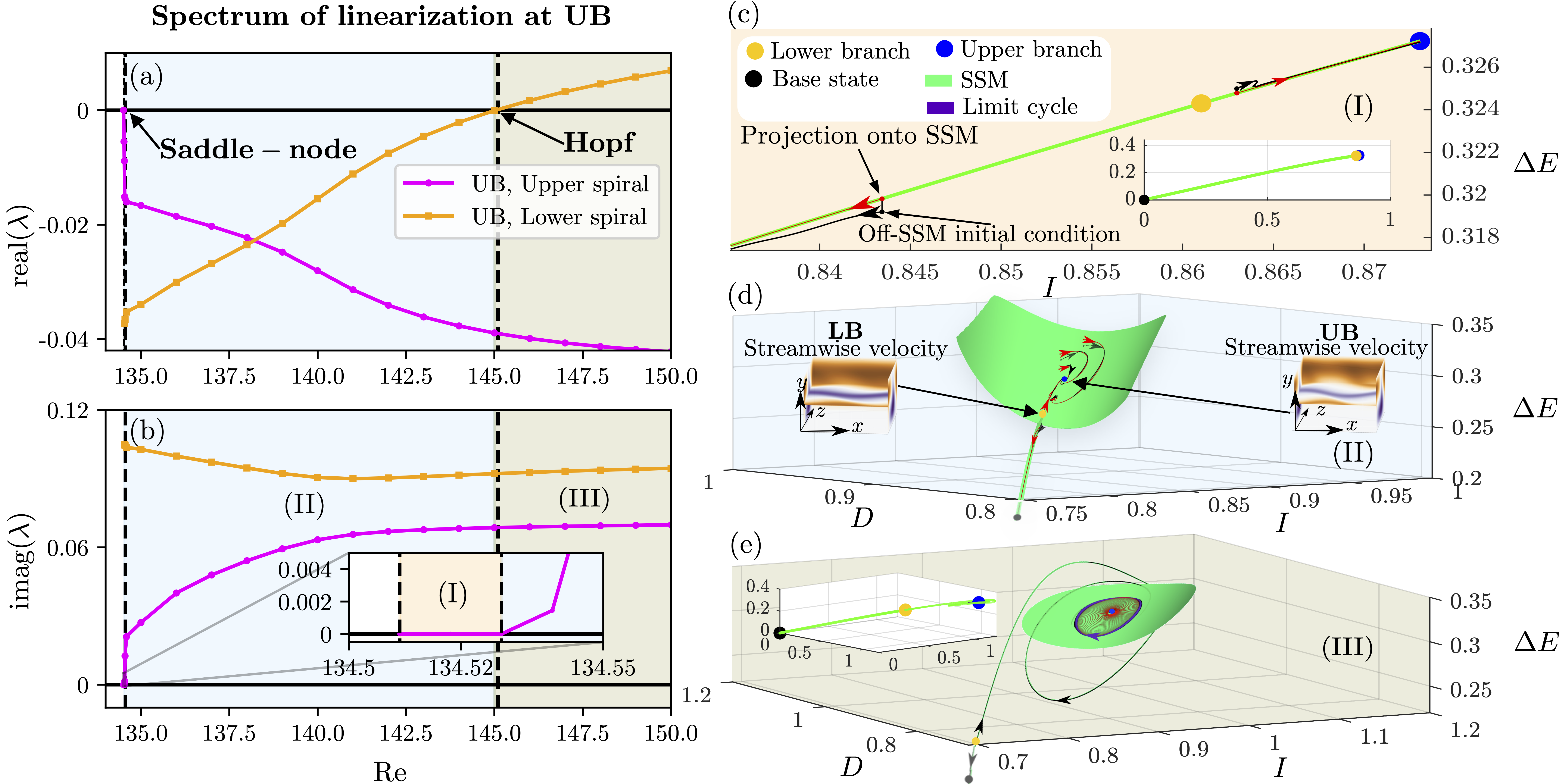}
    \caption{Spectral submanifolds (SSMs) with different dimensions. Panels (a) and (b) show the real and imaginary parts of the least stable eigenvalues $\lambda$ of the upper branch (UB) fixed point. The upper spiral (magenta curve) has a larger real part after the saddle-node bifurcation than the lower spiral (orange curve). In the shaded region marked by (I), the SSM is one-dimensional, while in regions (II) and (III), it is two-dimensional. In region (III), it also contains a stable limit cycle. Panels (c), (d), and (e) show the SSMs and the coherent states contained in them in the corresponding regions. In the insets of panel (d) the streamwise velocity of the lower- and upper branch fixed points is shown color coded. Panels (d) and (e) are not to scale. }
    \label{fig:2}
\end{figure*}

Within the Reynolds number interval $[134.5, 150]$, we focus our analysis either on the unstable manifold of the lower branch fixed point or the slowest stable SSM of the stable limit cycle bifurcating from the upper branch fixed point.  The unstable manifold of the lower branch is known to be one-dimensional for a wide range of Reynolds numbers \cite{wangLowerBranchCoherent2007}, making that fixed point an edge state \cite{skufcaEdgeChaosParallel2006, schneiderLaminarturbulentBoundaryPlane2008, avilaStreamwiseLocalizedSolutionsOnset2013}. As a result, we also find a one-dimensional connection between the lower branch fixed point and the stable base state. As for the upper branch, we distinguish three Reynolds number regimes, denoted by (I), (II), and (III)  in Fig. \ref{fig:2}, with different SSM geometries.

In region (I), close to the saddle-node bifurcation, the one-dimensional unstable manifold of the lower branch arrives necessarily tangent to the slowest SSM of the upper branch. This unstable manifold forms a heteroclinic connection between the two fixed points \cite{halcrowHeteroclinicConnectionsPlane2009}. {In this region, the slow SSM of the upper branch contains two heteroclinic connections: one between the lower- and upper branch fixed point and one between the lower- branch fixed point and the base state, as shown in Fig. \ref{fig:2}(c).} 

This alignment, however, breaks down at $\text{Re}=134.53$ in region (II). The slowest SSM of the upper branch becomes two-dimensional due to a collision of the least stable and second least stable eigenvalues, resulting in a pair of complex conjugate eigenvalues. After this secondary bifurcation, the upper branch becomes a stable spiral-type fixed point, preventing global one-dimensional SSM-based reductions. The heteroclinic connections, however, still exist in the two-dimensional slowest SSM of the upper branch fixed point, as shown in Fig. \ref{fig:2}(d).  

The upper branch undergoes yet another bifurcation at $\text{Re}=145$, losing stability in a Hopf bifurcation leading to a stable limit cycle in region (III). After this point, the heteroclinic connection between the lower- and upper branches disappears. Instead, an orbit connects the lower branch fixed point with the newly born limit cycle. As shown in Fig. \ref{fig:2}(e), the two-dimensional unstable manifold of the upper branch fixed point forms the slowest stable SSM of the limit cycle. The unstable manifold of the lower branch fixed point shoots into a higher-dimensional SSM of the limit cycle.

To parametrize these SSMs, we use the square root of energy input, $J$ in the one-dimensional region (I) and the the square roots of energy input and dissipation $(J,K)$ in the two-dimensional regions (II) and (III), all centered at the base state. We seek parametrizations of the form
\begin{align}
    \mathbf{u}(x,y,z,t) &= \mathbf{u}_{\text{base}} + \sum_{l=1}^M \mathbf{w}^{(1)}_l(x,y,z, \text{Re})J^l(t)\,\,\text{or} \label{eq:param1}    \\ 
    \mathbf{u}(x,y,z,t) &=\mathbf{u}_\text{base} + \sum_{\substack{l,m \\ l+m \leq M}} \mathbf{w}^{(2)}_{l,m}(x,y,z) J^l(t) K^m(t), \label{eq:param2}
\end{align}
%
{The coefficients $\mathbf{w}^{(1)}_l$ and  $\mathbf{w}^{(2)}_{l,m}$ are identified from data via regression, as described in the Supplemental Material \cite{supplementalmaterial}. The maximal polynomial order is  $M=6, 2, 5$ for the regions (I), (II) and (III), respectively}. {We allow for Re-dependence in polynomial coefficients only for the region (I) because this domain is small enough for a simple linear approximation of the Re-dependence to be justified. }

We initialize the training trajectories for the SSM-based models to lie approximately on the heteroclinic orbits between the ECSs. To this end, we start from initial conditions of the form $\mathbf{u}_{LB}\pm\varepsilon \mathbf{v}$, where $\varepsilon = O(10^{-5})$, and $\mathbf{v}$ is the unstable eigenvector of the lower branch fixed point. We compute the ECSs, their eigenvalues and eigenvectors with a Newton-Krylov solver \cite{viswanathRecurrentMotionsPlane2007}.
Each of these initial conditions lies on the heteroclinic orbits between the lower and either the upper branches or the base state, respectively. 
In region (III), we use a training trajectory starting from the unstable subspace of the upper branch fixed point. {This is to ensure that our training data accurately represents the behavior near the SSM.} 

Once we have identified the SSM geometry in the space of velocity measurements \cite{whitney}, we turn to modeling the reduced dynamics of parametrizing variables, considering  first cases (I) and (II).  {Using the training trajectories, which are sampled at integer time instants, we fit polynomial mappings of the form }
\begin{equation}\label{eq:rdynamicsI_II}
    J_{n+1} = R(J_{n}, \text{Re}) \quad \text{or }
    \begin{pmatrix} J_{n+1} \\ K_{n+1}\end{pmatrix}  = \begin{pmatrix} R_1(J_n, K_n) \\ R_2(J_n, K_n), \end{pmatrix}
\end{equation}
where we have parametric dependence for Re in case (I) that is able to capture the saddle-node bifurcation of the lower and upper branch fixed points \cite{coco}. Further details on the identified reduced discrete dynamical systems are given in the Supplemental Material \cite{supplementalmaterial}.

{In region (III), we were not able to construct a two-dimensional SSM model of the upper transition dynamics: the one-dimensional unstable manifold of the lower branch fixed point spirals onto a limit cycle and hence is not a differentiable manifold. One could still construct higher-dimensional SSMs that contain this orbit. Analyzing the spectrum of the limit cycle (see the Supplemental Material \cite{supplementalmaterial} for the spectrum), we find that including the second slowest mode in the underlying spectral subspace would lead to a four-dimensional model. Instead, we construct a local nonlinear model on the two-dimensional SSM  given by the unstable manifold of the upper branch fixed point. This manifold contains the upper branch fixed point, the limit cycle, and transitions between them. } In this case, we use the  data-driven extended normal form construction of  \texttt{SSMLearn}, which yields the continuous-time reduced model  
\begin{align} 
    \dot{\rho} &= 0.00171 \rho - 0.01063\rho^3 -0.01350 \rho^5 ,  \label{eq:rdynamicsIIIrho}  \\
    \dot{\theta} &=0.09290 - 0.01671\rho^2 + 0.01003 \rho^4.  \label{eq:rdynamicsIIItheta}
\end{align}
{The polar normal form variables $(\rho, \theta)$ are connected to $(J,K)$ via a nonlinear change of coordinates \cite{guckenheimerNonlinearOscillationsDynamical1983}, which is identified from data as we discuss in the Supplemental Material \cite{supplementalmaterial}. The data-driven model in (\ref{eq:rdynamicsIIIrho},\ref{eq:rdynamicsIIItheta}) takes the form of a Stuart-Landau equation \cite{ stuartNonlinearMechanicsHydrodynamic1958,landauFluidMechanics1959}. }

To illustrate the accuracy and predictive power of these discrete and continuous reduced-order models, we consider trajectories initialized away from the SSMs in all three cases. {We consider an ensemble of randomly chosen initial conditions whose $L^2$ distance from the unstable lower state is $10^{-2}$. Based on the SSM-reduced models we have obtained, we predict the time evolution of their $(J(t), K(t))$ coordinates from $(J(0), K(0))$} using Eqs. (\ref{eq:rdynamicsI_II}-\ref{eq:rdynamicsIIItheta}) and then extend these predictions into  the full phase space using the graph of the corresponding SSM in Eqs. (\ref{eq:param1},\ref{eq:param2}), with final evaluation of the model prediction  error ${|| \mathbf{u}(t) - \mathbf{u}(t)_{\text{true}}||_{L^2}}$. 

\begin{figure*}[t]
    \centering
    \includegraphics[width = \textwidth]{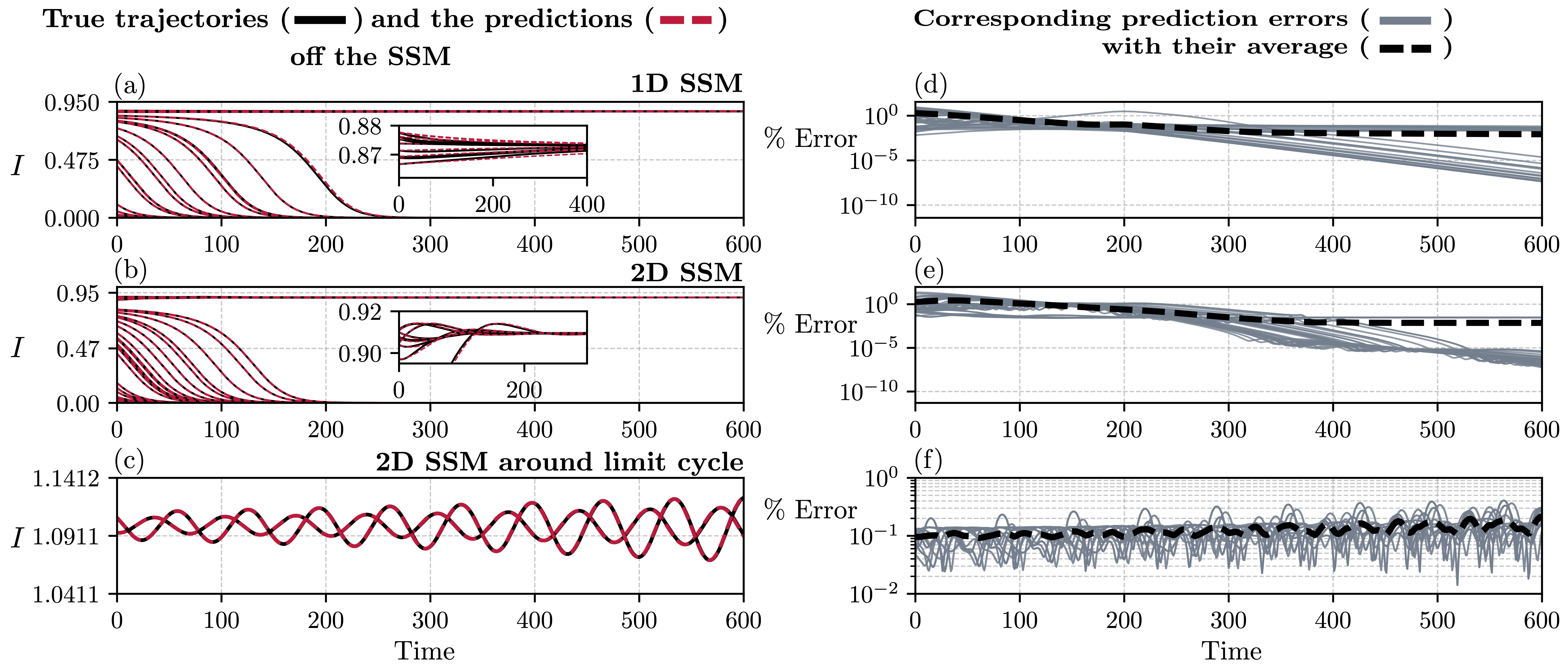}
    \caption{Predictions from the reduced-order models. In panels (a), (b) and (c) we show the predicted and true time evolutions of the variable $I$ for trajectories starting close to the SSM. The Reynolds number is 134.52 in (a), 135 in (b) and 146 in (c). In panels (d), (e) and (f) the prediction error $||\mathbf{u}(t) - \mathbf{u}_{\text{true}}(t)||/\text{max } ||\mathbf{u}_{\text{true}}(t)||$ is plotted in grey for the individual trajectories.  The black dashed curve shows the average of the error over the ensemble of 50 trajectories in panels (d), (e) and 20 trajectories in (f)). }
    \label{fig:3}
\end{figure*}

In panels (a,b,c) of Fig. \ref{fig:3}, we plot the time evolution of $I=J^2$ for a few members of the test ensemble. We found that the predictions based on the reduction to the SSM are qualitatively accurate as long as the trajectory does not start too far away from the SSM, where fast transients have a major influence on the dynamics. In panels (d,e,f) of Fig. \ref{fig:3}, we show the relative reconstruction error for the test ensemble. The error is normalized by the maximum value of the $L^{2}$ norm of the true trajectory. 
\section{Conclusions}
These results illustrate the power of SSM-based, data-driven reduced-order models for non-linearizable transition dynamics in a canonical nonlinear fluid flow that has defied prior attempts to derive such models. {We have demonstrated that SSM-reduced nonlinear models can simultaneously capture coexisting ECSs and heteroclinic transitions among them}, even in large distances from the stationary states serving as its anchor point. Importantly, our SSM-reduced models have also provided reliable predictions for transitions in open neighborhoods of the SSMs carrying them.

Consequently, more general tipping transitions \cite{ashwinTippingPointsOpen2012a, alkhayuonRateinducedTippingPeriodic2018}, i.e.,  transitions between steady states induced by changing parameters or noise, are also expected to be captured by an appropriate SSM-based reduction. Although we only considered the SSMs at low Reynolds in this shear flow configuration, a similar analysis is, in principle,  possible  for higher Reynolds numbers. At even higher Reynolds numbers,  the boundary of the domain of attraction of the base state (i.e., the stable manifold of the lower branch fixed point) becomes even more complicated, filled with unstable periodic orbits \cite{vanveenHomoclinicTangleEdge2011, kreilosPeriodicOrbitsOnset2012}. Reducing the dynamics to SSMs that are submanifolds of this edge offers hope for an SSM-based modeling of transition to turbulence. 

Limitations of the present approach include the a priori unknown size of the domain of validity of the reduced-order model in the full phase space and the lack of an appropriate SSM-based reduced-order model for upper transitions in domain (III).  Making progress on both of these challenges will require the construction of higher- (but still low-) dimensional invariant manifolds containing the upper transition orbits. Due to the simultaneous presence of both stable and unstable modes along such envisioned manifolds, the current SSM theory behind the \texttt{SSMLearn} algorithm needs technical extensions to accommodate both stable and unstable directions simultaneously. 

The codes and a sample of the data used for the analysis are available under the link \cite{ssmlearnCouette}, in the
form of a commented Matlab live script, as a part of the open-source toolbox SSMLearn \cite{cenedeseDatadrivenModelingPrediction2022}. The
complete data set is available from the authors upon request.

\begin{acknowledgments}
We are grateful to Jacob Page and Mingwu Li for useful discussions and their helpful comments. We acknowledge support from the Turbulent Superstructures Program (SPP1881) of the German National Science Foundation (DFG).
\end{acknowledgments}


\begin{thebibliography}{45}%
\makeatletter
\providecommand \@ifxundefined [1]{%
 \@ifx{#1\undefined}
}%
\providecommand \@ifnum [1]{%
 \ifnum #1\expandafter \@firstoftwo
 \else \expandafter \@secondoftwo
 \fi
}%
\providecommand \@ifx [1]{%
 \ifx #1\expandafter \@firstoftwo
 \else \expandafter \@secondoftwo
 \fi
}%
\providecommand \natexlab [1]{#1}%
\providecommand \enquote  [1]{``#1''}%
\providecommand \bibnamefont  [1]{#1}%
\providecommand \bibfnamefont [1]{#1}%
\providecommand \citenamefont [1]{#1}%
\providecommand \href@noop [0]{\@secondoftwo}%
\providecommand \href [0]{\begingroup \@sanitize@url \@href}%
\providecommand \@href[1]{\@@startlink{#1}\@@href}%
\providecommand \@@href[1]{\endgroup#1\@@endlink}%
\providecommand \@sanitize@url [0]{\catcode `\\12\catcode `\$12\catcode
  `\&12\catcode `\#12\catcode `\^12\catcode `\_12\catcode `\%12\relax}%
\providecommand \@@startlink[1]{}%
\providecommand \@@endlink[0]{}%
\providecommand \url  [0]{\begingroup\@sanitize@url \@url }%
\providecommand \@url [1]{\endgroup\@href {#1}{\urlprefix }}%
\providecommand \urlprefix  [0]{URL }%
\providecommand \Eprint [0]{\href }%
\providecommand \doibase [0]{https://doi.org/}%
\providecommand \selectlanguage [0]{\@gobble}%
\providecommand \bibinfo  [0]{\@secondoftwo}%
\providecommand \bibfield  [0]{\@secondoftwo}%
\providecommand \translation [1]{[#1]}%
\providecommand \BibitemOpen [0]{}%
\providecommand \bibitemStop [0]{}%
\providecommand \bibitemNoStop [0]{.\EOS\space}%
\providecommand \EOS [0]{\spacefactor3000\relax}%
\providecommand \BibitemShut  [1]{\csname bibitem#1\endcsname}%
\let\auto@bib@innerbib\@empty
\bibitem [{\citenamefont {Holmes}\ \emph {et~al.}(1996)\citenamefont {Holmes},
  \citenamefont {Lumley},\ and\ \citenamefont
  {Berkooz}}]{holmesTurbulenceCoherentStructures1996}%
  \BibitemOpen
  \bibfield  {author} {\bibinfo {author} {\bibfnamefont {P.}~\bibnamefont
  {Holmes}}, \bibinfo {author} {\bibfnamefont {J.~L.}\ \bibnamefont {Lumley}},\
  and\ \bibinfo {author} {\bibfnamefont {G.}~\bibnamefont {Berkooz}},\
  }\href@noop {} {\emph {\bibinfo {title} {Turbulence, {{Coherent Structures}},
  {{Dynamical Systems}} and {{Symmetry}}}}},\ Cambridge {{Monographs}} on
  {{Mechanics}}\ (\bibinfo  {publisher} {{Cambridge University Press}},\
  \bibinfo {address} {{Cambridge}},\ \bibinfo {year} {1996})\BibitemShut
  {NoStop}%
\bibitem [{\citenamefont {Schmid}(2022)}]{schmidDMDRev2022}%
  \BibitemOpen
  \bibfield  {author} {\bibinfo {author} {\bibfnamefont {P.~J.}\ \bibnamefont
  {Schmid}},\ }\bibfield  {title} {\bibinfo {title} {Dynamic mode decomposition
  and its variants},\ }\href
  {https://doi.org/10.1146/annurev-fluid-030121-015835} {\bibfield  {journal}
  {\bibinfo  {journal} {Ann. Rev. Fluid Mech.}\ }\textbf {\bibinfo {volume}
  {54}},\ \bibinfo {pages} {225} (\bibinfo {year} {2022})}\BibitemShut
  {NoStop}%
\bibitem [{\citenamefont {Page}\ and\ \citenamefont
  {Kerswell}(2019)}]{pageKoopmanModeExpansions2019}%
  \BibitemOpen
  \bibfield  {author} {\bibinfo {author} {\bibfnamefont {J.}~\bibnamefont
  {Page}}\ and\ \bibinfo {author} {\bibfnamefont {R.~R.}\ \bibnamefont
  {Kerswell}},\ }\bibfield  {title} {\bibinfo {title} {Koopman mode expansions
  between simple invariant solutions},\ }\href@noop {} {\bibfield  {journal}
  {\bibinfo  {journal} {J. Fluid Mech.}\ }\textbf {\bibinfo {volume} {879}},\
  \bibinfo {pages} {1} (\bibinfo {year} {2019})}\BibitemShut {NoStop}%
\bibitem [{\citenamefont {Kawahara}\ \emph {et~al.}(2012)\citenamefont
  {Kawahara}, \citenamefont {Uhlmann},\ and\ \citenamefont {van
  Veen}}]{kawahara_significance_2012}%
  \BibitemOpen
  \bibfield  {author} {\bibinfo {author} {\bibfnamefont {G.}~\bibnamefont
  {Kawahara}}, \bibinfo {author} {\bibfnamefont {M.}~\bibnamefont {Uhlmann}},\
  and\ \bibinfo {author} {\bibfnamefont {L.}~\bibnamefont {van Veen}},\
  }\bibfield  {title} {\bibinfo {title} {The {Significance} of {Simple}
  {Invariant} {Solutions} in {Turbulent} {Flows}},\ }\href
  {https://doi.org/10.1146/annurev-fluid-120710-101228} {\bibfield  {journal}
  {\bibinfo  {journal} {Annu. Rev. Fluid Mech.}\ }\textbf {\bibinfo {volume}
  {44}},\ \bibinfo {pages} {203} (\bibinfo {year} {2012})}\BibitemShut
  {NoStop}%
\bibitem [{\citenamefont {Brunton}\ \emph {et~al.}(2020)\citenamefont
  {Brunton}, \citenamefont {Noack},\ and\ \citenamefont
  {Koumoutsakos}}]{bruntonRev2020}%
  \BibitemOpen
  \bibfield  {author} {\bibinfo {author} {\bibfnamefont {S.}~\bibnamefont
  {Brunton}}, \bibinfo {author} {\bibfnamefont {B.}~\bibnamefont {Noack}},\
  and\ \bibinfo {author} {\bibfnamefont {P.}~\bibnamefont {Koumoutsakos}},\
  }\bibfield  {title} {\bibinfo {title} {Machine learning for fluid
  mechanics},\ }\href@noop {} {\bibfield  {journal} {\bibinfo  {journal} {Ann.
  Rev. Fluid Mech.}\ }\textbf {\bibinfo {volume} {52}},\ \bibinfo {pages} {477}
  (\bibinfo {year} {2020})}\BibitemShut {NoStop}%
\bibitem [{\citenamefont {Haller}\ and\ \citenamefont
  {Ponsioen}(2016)}]{hallerNonlinearNormalModes2016a}%
  \BibitemOpen
  \bibfield  {author} {\bibinfo {author} {\bibfnamefont {G.}~\bibnamefont
  {Haller}}\ and\ \bibinfo {author} {\bibfnamefont {S.}~\bibnamefont
  {Ponsioen}},\ }\bibfield  {title} {\bibinfo {title} {Nonlinear normal modes
  and spectral submanifolds: Existence, uniqueness and use in model
  reduction},\ }\href@noop {} {\bibfield  {journal} {\bibinfo  {journal}
  {Nonlinear Dyn}\ }\textbf {\bibinfo {volume} {86}},\ \bibinfo {pages} {1493}
  (\bibinfo {year} {2016})}\BibitemShut {NoStop}%
  \bibitem [{\citenamefont {Kogelbauer}\ and\ \citenamefont
  {Haller}(2018)}]{kogelbauerRigorousModelReduction2018}%
  \BibitemOpen
  \bibfield  {author} {\bibinfo {author} {\bibfnamefont {F.}~\bibnamefont
  {Kogelbauer}}\ and\ \bibinfo {author} {\bibfnamefont {G.}~\bibnamefont
  {Haller}},\ }\bibfield  {title} { {\bibinfo {title} {Rigorous {{Model
  Reduction}} for a {{Damped-Forced Nonlinear Beam Model}}: {{An
  Infinite-Dimensional Analysis}}},\ }}\href@noop {} {\bibfield  {journal}
  {\bibinfo  {journal} {J Nonlinear Sci}\ }\textbf {\bibinfo {volume} {28}},\
  \bibinfo {pages} {1109} (\bibinfo {year} {2018})}\BibitemShut {NoStop}%
\bibitem [{\citenamefont {Cabre}\ \emph {et~al.}(2003)\citenamefont {Cabre},
  \citenamefont {Fontich},\ and\ \citenamefont {{de la
  Llave}}}]{cabreParameterizationMethodInvariant2003a}%
  \BibitemOpen
  \bibfield  {author} {\bibinfo {author} {\bibfnamefont {X.}~\bibnamefont
  {Cabre}}, \bibinfo {author} {\bibfnamefont {E.}~\bibnamefont {Fontich}},\
  and\ \bibinfo {author} {\bibfnamefont {R.}~\bibnamefont {{de la Llave}}},\
  }\bibfield  {title} {\bibinfo {title} {The parameterization method for
  invariant manifolds {{I}}: {{Manifolds}} associated to non-resonant
  subspaces},\ }\href@noop {} {\bibfield  {journal} {\bibinfo  {journal}
  {Indiana Univ. Math. J.}\ }\textbf {\bibinfo {volume} {52}},\ \bibinfo
  {pages} {283} (\bibinfo {year} {2003})}\BibitemShut {NoStop}%
\bibitem [{\citenamefont {Haro}\ and\ \citenamefont {de~la
  Llave}(2006)}]{haro2006}%
  \BibitemOpen
  \bibfield  {author} {\bibinfo {author} {\bibfnamefont {A.}~\bibnamefont
  {Haro}}\ and\ \bibinfo {author} {\bibfnamefont {R.}~\bibnamefont {de~la
  Llave}},\ }\bibfield  {title} {\bibinfo {title} {A parameterization method
  for the computation of invariant tori and their whiskers in quasi-periodic
  maps: Rigorous results},\ }\href@noop {} {\bibfield  {journal} {\bibinfo
  {journal} {J. Diff. Eqs.}\ }\textbf {\bibinfo {volume} {228}},\ \bibinfo
  {pages} {530} (\bibinfo {year} {2006})}\BibitemShut {NoStop}%
  \bibitem [{\citenamefont {Kaszás}, \ \citenamefont
  {Cenedese and Haller}()}]{supplementalmaterial}%
  \BibitemOpen
  \bibfield  {author} {\bibinfo {author} {\bibfnamefont {B.}~\bibnamefont
  {Kaszás}}, \ \bibinfo {author} {\bibfnamefont {M.}~\bibnamefont
  {Cenedese}}\ and \ \bibinfo {author} {\bibfnamefont {G.}~\bibnamefont
  {Haller}} }\href@noop {} { {\bibinfo {title} Dynamics-based machine learning of transitions in Couette flow: Supplemental Material}},\ \bibinfo {note} {{\tt {
  Link will be inserted by publisher}}}\BibitemShut {NoStop}%
\bibitem [{\citenamefont {Cenedese}\ \emph
  {et~al.}(2022{\natexlab{a}})\citenamefont {Cenedese}, \citenamefont
  {Ax{\aa}s}, \citenamefont {B{\"a}uerlein}, \citenamefont {Avila},\ and\
  \citenamefont {Haller}}]{cenedeseDatadrivenModelingPrediction2022}%
  \BibitemOpen
  \bibfield  {author} {\bibinfo {author} {\bibfnamefont {M.}~\bibnamefont
  {Cenedese}}, \bibinfo {author} {\bibfnamefont {J.}~\bibnamefont {Ax{\aa}s}},
  \bibinfo {author} {\bibfnamefont {B.}~\bibnamefont {B{\"a}uerlein}}, \bibinfo
  {author} {\bibfnamefont {K.}~\bibnamefont {Avila}},\ and\ \bibinfo {author}
  {\bibfnamefont {G.}~\bibnamefont {Haller}},\ }\bibfield  {title} {\bibinfo
  {title} {Data-driven modeling and prediction of non-linearizable dynamics via
  spectral submanifolds},\ }\href@noop {} {\bibfield  {journal} {\bibinfo
  {journal} {Nature Commun}\ }\textbf {\bibinfo {volume} {13}},\ \bibinfo
  {pages} {872} (\bibinfo {year} {2022}{\natexlab{a}})}\BibitemShut {NoStop}%
\bibitem [{\citenamefont {Jain}\ and\ \citenamefont
  {Haller}(2022)}]{jainHowComputeInvariant2022}%
  \BibitemOpen
  \bibfield  {author} {\bibinfo {author} {\bibfnamefont {S.}~\bibnamefont
  {Jain}}\ and\ \bibinfo {author} {\bibfnamefont {G.}~\bibnamefont {Haller}},\
  }\bibfield  {title} {\bibinfo {title} {How to compute invariant manifolds and
  their reduced dynamics in high-dimensional finite element models},\
  }\href@noop {} {\bibfield  {journal} {\bibinfo  {journal} {Nonlinear Dyn}\
  }\textbf {\bibinfo {volume} {107}},\ \bibinfo {pages} {1417} (\bibinfo {year}
  {2022})}\BibitemShut {NoStop}%
\bibitem [{\citenamefont {Li}\ \emph {et~al.}(2020)\citenamefont {Li},
  \citenamefont {Jain},\ and\ \citenamefont {Haller}}]{li2022}%
  \BibitemOpen
  \bibfield  {author} {\bibinfo {author} {\bibfnamefont {M.}~\bibnamefont
  {Li}}, \bibinfo {author} {\bibfnamefont {S.}~\bibnamefont {Jain}},\ and\
  \bibinfo {author} {\bibfnamefont {G.}~\bibnamefont {Haller}},\ }\bibfield
  {title} {\bibinfo {title} {Nonlinear analysis of forced mechanical systems
  with internal resonance using spectral submanifolds -- {P}art I: {P}eriodic
  response and forced response curve},\ }\href@noop {} \href@noop {} {\bibfield  {journal} {\bibinfo  {journal}
  {Nonlinear Dyn},} (\bibinfo {year} {2022})}\BibitemShut {NoStop}%
\bibitem [{\citenamefont {Cenedese}\ \emph
  {et~al.}(2022{\natexlab{b}})\citenamefont {Cenedese}, \citenamefont
  {Ax{\aa}s}, \citenamefont {Yang}, \citenamefont {Eriten},\ and\ \citenamefont
  {Haller}}]{cenedeseDatadrivenModelingPrediction2022b}%
  \BibitemOpen
  \bibfield  {author} {\bibinfo {author} {\bibfnamefont {M.}~\bibnamefont
  {Cenedese}}, \bibinfo {author} {\bibfnamefont {J.}~\bibnamefont {Ax{\aa}s}},
  \bibinfo {author} {\bibfnamefont {H.}~\bibnamefont {Yang}}, \bibinfo {author}
  {\bibfnamefont {M.}~\bibnamefont {Eriten}},\ and\ \bibinfo {author}
  {\bibfnamefont {G.}~\bibnamefont {Haller}},\ }\bibfield  {title} {\bibinfo
  {title} {Data-driven nonlinear model reduction to spectral submanifolds in
  mechanical systems},\ }\href@noop {} {\bibfield  {journal} {\bibinfo
  {journal} {Phil. Trans. R. Soc. A}\ }\textbf {\bibinfo {volume} {380}},\
  \bibinfo {pages} {20210194} (\bibinfo {year}
  {2022}{\natexlab{b}})}\BibitemShut {NoStop}%
\bibitem [{\citenamefont {Ax{\aa}s}\ \emph {et~al.}(2022)\citenamefont
  {Ax{\aa}s}, \citenamefont {Cenedese},\ and\ \citenamefont
  {Haller}}]{axas2022}%
  \BibitemOpen
  \bibfield  {author} {\bibinfo {author} {\bibfnamefont {J.}~\bibnamefont
  {Ax{\aa}s}}, \bibinfo {author} {\bibfnamefont {M.}~\bibnamefont {Cenedese}},\
  and\ \bibinfo {author} {\bibfnamefont {G.}~\bibnamefont {Haller}},\
  }\bibfield  {title} {\bibinfo {title} {Fast data-driven model reduction for
  nonlinear dynamical systems},\ }\href@noop {} {\bibfield  {journal} {\bibinfo
   {journal} {arXiv:2204.14169}\ } (\bibinfo {year} {2022})}\BibitemShut
  {NoStop}%
\bibitem [{\citenamefont {Lenton}\ \emph {et~al.}(2008)\citenamefont {Lenton},
  \citenamefont {Held}, \citenamefont {Kriegler}, \citenamefont {Hall},
  \citenamefont {Lucht}, \citenamefont {Rahmstorf},\ and\ \citenamefont
  {Schellnhuber}}]{lenton_tipping2008}%
  \BibitemOpen
  \bibfield  {author} {\bibinfo {author} {\bibfnamefont {T.~M.}\ \bibnamefont
  {Lenton}}, \bibinfo {author} {\bibfnamefont {H.}~\bibnamefont {Held}},
  \bibinfo {author} {\bibfnamefont {E.}~\bibnamefont {Kriegler}}, \bibinfo
  {author} {\bibfnamefont {J.~W.}\ \bibnamefont {Hall}}, \bibinfo {author}
  {\bibfnamefont {W.}~\bibnamefont {Lucht}}, \bibinfo {author} {\bibfnamefont
  {S.}~\bibnamefont {Rahmstorf}},\ and\ \bibinfo {author} {\bibfnamefont
  {H.~J.}\ \bibnamefont {Schellnhuber}},\ }\bibfield  {title} {\bibinfo {title}
  {Tipping elements in the Earth's climate system},\ }\href
  {https://doi.org/10.1073/pnas.0705414105} {\bibfield  {journal} {\bibinfo
  {journal} {Proc. Natl. Acad. Sci. U.S.A.,}\ }\textbf {\bibinfo {volume}
  {105}},\ \bibinfo {pages} {1786} (\bibinfo {year} {2008})}\BibitemShut
  {NoStop}%
\bibitem [{\citenamefont {Waleffe}(2001)}]{waleffeExactCoherentStructures2001}%
  \BibitemOpen
  \bibfield  {author} {\bibinfo {author} {\bibfnamefont {F.}~\bibnamefont
  {Waleffe}},\ }\bibfield  {title} {\bibinfo {title} {Exact coherent structures
  in channel flow},\ }\href@noop {} {\bibfield  {journal} {\bibinfo  {journal}
  {J. Fluid Mech.}\ }\textbf {\bibinfo {volume} {435}},\ \bibinfo {pages} {93}
  (\bibinfo {year} {2001})}\BibitemShut {NoStop}%
\bibitem [{\citenamefont {Gibson}\ \emph {et~al.}(2008)\citenamefont {Gibson},
  \citenamefont {Halcrow},\ and\ \citenamefont
  {Cvitanovi{\'c}}}]{gibsonVisualizingGeometryState2008}%
  \BibitemOpen
  \bibfield  {author} {\bibinfo {author} {\bibfnamefont {J.~F.}\ \bibnamefont
  {Gibson}}, \bibinfo {author} {\bibfnamefont {J.}~\bibnamefont {Halcrow}},\
  and\ \bibinfo {author} {\bibfnamefont {P.}~\bibnamefont {Cvitanovi{\'c}}},\
  }\bibfield  {title} {\bibinfo {title} {Visualizing the geometry of state
  space in plane {Couette} flow},\ }\href@noop {} {\bibfield  {journal}
  {\bibinfo  {journal} {J. Fluid Mech.}\ }\textbf {\bibinfo {volume} {611}},\
  \bibinfo {pages} {107} (\bibinfo {year} {2008})}\BibitemShut {NoStop}%
\bibitem [{\citenamefont {Fujimura}(1997)}]{fujimuraCenter1997}%
  \BibitemOpen
  \bibfield  {author} {\bibinfo {author} {\bibfnamefont {K.}~\bibnamefont
  {Fujimura}},\ }\bibfield  {title} {\bibinfo {title} {Centre manifold
  reduction and the {S}tuart-{L}andau equation for fluid motions},\ }\href@noop
  {} {\bibfield  {journal} {\bibinfo  {journal} {Proc. R. Soc. Lond. A.}\
  }\textbf {\bibinfo {volume} {453}},\ \bibinfo {pages} {181} (\bibinfo {year}
  {1997})}\BibitemShut {NoStop}%
\bibitem [{\citenamefont {Carini}\ \emph {et~al.}(2015)\citenamefont {Carini},
  \citenamefont {Auteri},\ and\ \citenamefont
  {Giannetti}}]{cariniCentremanifoldReductionBifurcating2015}%
  \BibitemOpen
  \bibfield  {author} {\bibinfo {author} {\bibfnamefont {M.}~\bibnamefont
  {Carini}}, \bibinfo {author} {\bibfnamefont {F.}~\bibnamefont {Auteri}},\
  and\ \bibinfo {author} {\bibfnamefont {F.}~\bibnamefont {Giannetti}},\
  }\bibfield  {title} {\bibinfo {title} {Centre-manifold reduction of
  bifurcating flows},\ }\href@noop {} {\bibfield  {journal} {\bibinfo
  {journal} {J. Fluid Mech.}\ }\textbf {\bibinfo {volume} {767}},\ \bibinfo
  {pages} {109} (\bibinfo {year} {2015})}\BibitemShut {NoStop}%
\bibitem [{\citenamefont {Budanur}\ and\ \citenamefont
  {Cvitanovi{\'c}}(2017)}]{budanurUnstableManifoldsRelative2017}%
  \BibitemOpen
  \bibfield  {author} {\bibinfo {author} {\bibfnamefont {N.~B.}\ \bibnamefont
  {Budanur}}\ and\ \bibinfo {author} {\bibfnamefont {P.}~\bibnamefont
  {Cvitanovi{\'c}}},\ }\bibfield  {title} {\bibinfo {title} {Unstable
  {{Manifolds}} of {{Relative Periodic Orbits}} in the {{Symmetry-Reduced State
  Space}} of the {{Kuramoto}}\textendash{{Sivashinsky System}}},\ }\href@noop
  {} {\bibfield  {journal} {\bibinfo  {journal} {J Stat Phys}\ }\textbf
  {\bibinfo {volume} {167}},\ \bibinfo {pages} {636} (\bibinfo {year}
  {2017})}\BibitemShut {NoStop}%
\bibitem [{\citenamefont {Budanur}\ and\ \citenamefont
  {Hof}(2017)}]{budanurHeteroclinicPathSpatially2017a}%
  \BibitemOpen
  \bibfield  {author} {\bibinfo {author} {\bibfnamefont {N.~B.}\ \bibnamefont
  {Budanur}}\ and\ \bibinfo {author} {\bibfnamefont {B.}~\bibnamefont {Hof}},\
  }\bibfield  {title} {\bibinfo {title} {Heteroclinic path to spatially
  localized chaos in pipe flow},\ }\href@noop {} {\bibfield  {journal}
  {\bibinfo  {journal} {J. Fluid Mech.}\ }\textbf {\bibinfo {volume} {827}},\
  \bibinfo {pages} {R1} (\bibinfo {year} {2017})}\BibitemShut {NoStop}%
\bibitem [{\citenamefont {Farano}\ \emph {et~al.}(2019)\citenamefont {Farano},
  \citenamefont {Cherubini}, \citenamefont {Robinet}, \citenamefont
  {De~Palma},\ and\ \citenamefont
  {Schneider}}]{faranoComputingHeteroclinicOrbits2019}%
  \BibitemOpen
  \bibfield  {author} {\bibinfo {author} {\bibfnamefont {M.}~\bibnamefont
  {Farano}}, \bibinfo {author} {\bibfnamefont {S.}~\bibnamefont {Cherubini}},
  \bibinfo {author} {\bibfnamefont {J.-C.}\ \bibnamefont {Robinet}}, \bibinfo
  {author} {\bibfnamefont {P.}~\bibnamefont {De~Palma}},\ and\ \bibinfo
  {author} {\bibfnamefont {T.~M.}\ \bibnamefont {Schneider}},\ }\bibfield
  {title} {\bibinfo {title} {Computing heteroclinic orbits using adjoint-based
  methods},\ }\href@noop {} {\bibfield  {journal} {\bibinfo  {journal} {J.
  Fluid Mech.}\ }\textbf {\bibinfo {volume} {858}},\ \bibinfo {pages} {R3}
  (\bibinfo {year} {2019})}\BibitemShut {NoStop}%
\bibitem [{\citenamefont {Suri}\ \emph {et~al.}(2017)\citenamefont {Suri},
  \citenamefont {Tithof}, \citenamefont {Grigoriev},\ and\ \citenamefont
  {Schatz}}]{suriForecastingFluidFlows2017}%
  \BibitemOpen
  \bibfield  {author} {\bibinfo {author} {\bibfnamefont {B.}~\bibnamefont
  {Suri}}, \bibinfo {author} {\bibfnamefont {J.}~\bibnamefont {Tithof}},
  \bibinfo {author} {\bibfnamefont {R.~O.}\ \bibnamefont {Grigoriev}},\ and\
  \bibinfo {author} {\bibfnamefont {M.~F.}\ \bibnamefont {Schatz}},\ }\bibfield
   {title} {\bibinfo {title} {Forecasting {{Fluid Flows Using}} the
  {{Geometry}} of {{Turbulence}}},\ }\href@noop {} {\bibfield  {journal}
  {\bibinfo  {journal} {Phys. Rev. Lett.}\ }\textbf {\bibinfo {volume} {118}},\
  \bibinfo {pages} {114501} (\bibinfo {year} {2017})}\BibitemShut {NoStop}%
\bibitem [{\citenamefont {Loiseau}\ \emph {et~al.}(2020)\citenamefont
  {Loiseau}, \citenamefont {Brunton},\ and\ \citenamefont
  {Noack}}]{loiseauPODmanifold2020}%
  \BibitemOpen
  \bibfield  {author} {\bibinfo {author} {\bibfnamefont {J.-C.}\ \bibnamefont
  {Loiseau}}, \bibinfo {author} {\bibfnamefont {S.}~\bibnamefont {Brunton}},\
  and\ \bibinfo {author} {\bibfnamefont {B.}~\bibnamefont {Noack}},\ }\bibinfo
  {title} {From the POD-Galerkin method to sparse manifold models},\ in\
  \href@noop {} {\emph {\bibinfo {booktitle} {Model Order Reduction, Volume 3:
  Applications}}},\ \bibinfo {editor} {edited by\ \bibinfo {editor}
  {\bibfnamefont {P.}~\bibnamefont {Benner}}, \bibinfo {editor} {\bibfnamefont
  {S.}~\bibnamefont {Grivet-Talocia}}, \bibinfo {editor} {\bibfnamefont
  {A.}~\bibnamefont {Quarteroni}}, \bibinfo {editor} {\bibfnamefont
  {G.}~\bibnamefont {Rozza}}, \bibinfo {editor} {\bibfnamefont
  {W.}~\bibnamefont {Schilders}},\ and\ \bibinfo {editor} {\bibfnamefont
  {L.}~\bibnamefont {Silveira}}}\ (\bibinfo  {publisher} {De Gruyter, Berlin},\
  \bibinfo {year} {2020})\ pp.\ \bibinfo {pages} {279--320}\BibitemShut
  {NoStop}%
\bibitem [{\citenamefont {Rowley}\ \emph {et~al.}(2009)\citenamefont {Rowley},
  \citenamefont {Mezi{\'c}}, \citenamefont {Bagheri}, \citenamefont
  {Schlatter},\ and\ \citenamefont
  {Henningson}}]{rowleySpectralAnalysisNonlinear2009}%
  \BibitemOpen
  \bibfield  {author} {\bibinfo {author} {\bibfnamefont {C.~W.}\ \bibnamefont
  {Rowley}}, \bibinfo {author} {\bibfnamefont {I.}~\bibnamefont {Mezi{\'c}}},
  \bibinfo {author} {\bibfnamefont {S.}~\bibnamefont {Bagheri}}, \bibinfo
  {author} {\bibfnamefont {P.}~\bibnamefont {Schlatter}},\ and\ \bibinfo
  {author} {\bibfnamefont {D.~S.}\ \bibnamefont {Henningson}},\ }\bibfield
  {title} {\bibinfo {title} {Spectral analysis of nonlinear flows},\
  }\href@noop {} {\bibfield  {journal} {\bibinfo  {journal} {J. Fluid Mech.}\
  }\textbf {\bibinfo {volume} {641}},\ \bibinfo {pages} {115} (\bibinfo {year}
  {2009})}\BibitemShut {NoStop}%
\bibitem [{\citenamefont {Hastie}\ \emph {et~al.}(2009)\citenamefont {Hastie},
  \citenamefont {Tibshirani},\ and\ \citenamefont
  {Friedman}}]{hastie2009elements}%
  \BibitemOpen
  \bibfield  {author} {\bibinfo {author} {\bibfnamefont {T.}~\bibnamefont
  {Hastie}}, \bibinfo {author} {\bibfnamefont {R.}~\bibnamefont {Tibshirani}},\
  and\ \bibinfo {author} {\bibfnamefont {J.}~\bibnamefont {Friedman}},\ }\href
  {https://books.google.ch/books?id=tVIjmNS3Ob8C} {\emph {\bibinfo {title} {The
  Elements of Statistical Learning: Data Mining, Inference, and Prediction,
  Second Edition}}},\ Springer Series in Statistics\ (\bibinfo  {publisher}
  {Springer New York},\ \bibinfo {year} {2009})\BibitemShut {NoStop}%
\bibitem [{\citenamefont {Kaszás}\ and\ \citenamefont
  {Cenedese}()}]{ssmlearnCouette}%
  \BibitemOpen
  \bibfield  {author} {\bibinfo {author} {\bibfnamefont {B.}~\bibnamefont
  {Kaszás}}\ and\ \bibinfo {author} {\bibfnamefont {M.}~\bibnamefont
  {Cenedese}},\ }\href@noop {} { {\bibinfo {title} {Analysis of {C}ouette
  Flow Regimes via {{\tt {SSMLearn}}}}}}. The data and the code is available under the repository, \ \bibinfo {note} {{\tt {\url{
  https://github.com/haller-group/SSMLearn/tree/main/examples/couetteflow}}}} (2022)\BibitemShut {NoStop}%
  \bibitem [{\citenamefont {Orr}(1907)}]{orrStabilityInstabilitySteady1907}%
  \BibitemOpen
  \bibfield  {author} {\bibinfo {author} {\bibfnamefont {W.~M.}\ \bibnamefont
  {Orr}},\ }\bibfield  {title} { {\bibinfo {title} {The {{Stability}} or
  {{Instability}} of the {{Steady Motions}} of a {{Perfect Liquid}} and of a
  {{Viscous Liquid}}. {{Part II}}: {{A Viscous Liquid}}},\ }}\href@noop {}
  {\bibfield  {journal} {\bibinfo  {journal} {Proceedings of the Royal Irish
  Academy. Section A: Mathematical and Physical Sciences}\ }\textbf {\bibinfo
  {volume} {27}},\ \bibinfo {pages} {69} (\bibinfo {year} {1907})}\BibitemShut
  {NoStop}%
\bibitem [{\citenamefont {Gibson}\ \emph {et~al.}(2009)\citenamefont {Gibson},
  \citenamefont {Halcrow},\ and\ \citenamefont
  {Cvitanovi{\'c}}}]{gibsonEquilibriumTravellingwaveSolutions2009}%
  \BibitemOpen
  \bibfield  {author} {\bibinfo {author} {\bibfnamefont {J.~F.}\ \bibnamefont
  {Gibson}}, \bibinfo {author} {\bibfnamefont {J.}~\bibnamefont {Halcrow}},\
  and\ \bibinfo {author} {\bibfnamefont {P.}~\bibnamefont {Cvitanovi{\'c}}},\
  }\bibfield  {title} {\bibinfo {title} {Equilibrium and travelling-wave
  solutions of plane {Couette} flow},\ }\href@noop {} {\bibfield  {journal}
  {\bibinfo  {journal} {J. Fluid Mech.}\ }\textbf {\bibinfo {volume} {638}},\
  \bibinfo {pages} {243} (\bibinfo {year} {2009})}\BibitemShut {NoStop}%
\bibitem [{\citenamefont
  {Nagata}(1990)}]{nagataThreedimensionalFiniteamplitudeSolutions1990}%
  \BibitemOpen
  \bibfield  {author} {\bibinfo {author} {\bibfnamefont {M.}~\bibnamefont
  {Nagata}},\ }\bibfield  {title} {\bibinfo {title} {Three-dimensional
  finite-amplitude solutions in plane {{Couette}} flow: Bifurcation from
  infinity},\ }\href@noop {} {\bibfield  {journal} {\bibinfo  {journal} {J.
  Fluid Mech.}\ }\textbf {\bibinfo {volume} {217}},\ \bibinfo {pages} {519}
  (\bibinfo {year} {1990})}\BibitemShut {NoStop}%
\bibitem [{\citenamefont {Gibson}\ \emph {et~al.}(2019)\citenamefont {Gibson},
  \citenamefont {Reetz}, \citenamefont {Azimi}, \citenamefont {Ferraro},
  \citenamefont {Kreilos}, \citenamefont {Schrobsdorff}, \citenamefont
  {Farano}, \citenamefont {Yesil}, \citenamefont {Schütz}, \citenamefont
  {Culpo},\ and\ \citenamefont {Schneider}}]{channelflow}%
  \BibitemOpen
  \bibfield  {author} {\bibinfo {author} {\bibfnamefont {J.~F.}\ \bibnamefont
  {Gibson}}, \bibinfo {author} {\bibfnamefont {F.}~\bibnamefont {Reetz}},
  \bibinfo {author} {\bibfnamefont {S.}~\bibnamefont {Azimi}}, \bibinfo
  {author} {\bibfnamefont {A.}~\bibnamefont {Ferraro}}, \bibinfo {author}
  {\bibfnamefont {T.}~\bibnamefont {Kreilos}}, \bibinfo {author} {\bibfnamefont
  {H.}~\bibnamefont {Schrobsdorff}}, \bibinfo {author} {\bibfnamefont
  {M.}~\bibnamefont {Farano}}, \bibinfo {author} {\bibfnamefont {A.~F.}\
  \bibnamefont {Yesil}}, \bibinfo {author} {\bibfnamefont {S.~S.}\ \bibnamefont
  {Schütz}}, \bibinfo {author} {\bibfnamefont {M.}~\bibnamefont {Culpo}},\
  and\ \bibinfo {author} {\bibfnamefont {T.~M.}\ \bibnamefont {Schneider}},\
  }\href {https://channelflow.ch} {\emph {\bibinfo {title} {Channelflow
  2.0}}},\ \bibinfo {type} {Tech. Rep.}\ (\bibinfo  {institution} {manuscript
  in preparation},\ \bibinfo {year} {2019})\ \bibinfo {note} {{\tt
  {channelflow.ch}}}\BibitemShut {NoStop}%
\bibitem [{\citenamefont {Waleffe}(2003)}]{waleffeHomotopyExactCoherent2003}%
  \BibitemOpen
  \bibfield  {author} {\bibinfo {author} {\bibfnamefont {F.}~\bibnamefont
  {Waleffe}},\ }\bibfield  {title} {\bibinfo {title} {Homotopy of exact
  coherent structures in plane shear flows},\ }\href@noop {} {\bibfield
  {journal} {\bibinfo  {journal} {Phys. Fluids}\ }\textbf {\bibinfo {volume}
  {15}},\ \bibinfo {pages} {1517} (\bibinfo {year} {2003})}\BibitemShut
  {NoStop}%
\bibitem [{\citenamefont {Kawahara}\ and\ \citenamefont
  {Kida}(2001)}]{kawaharaPeriodicMotionEmbedded2001}%
  \BibitemOpen
  \bibfield  {author} {\bibinfo {author} {\bibfnamefont {G.}~\bibnamefont
  {Kawahara}}\ and\ \bibinfo {author} {\bibfnamefont {S.}~\bibnamefont
  {Kida}},\ }\bibfield  {title} {\bibinfo {title} {Periodic motion embedded in
  plane {{Couette}} turbulence: Regeneration cycle and burst},\ }\href@noop {}
  {\bibfield  {journal} {\bibinfo  {journal} {J. Fluid Mech.}\ }\textbf
  {\bibinfo {volume} {449}},\ \bibinfo {pages} {291} (\bibinfo {year}
  {2001})}\BibitemShut {NoStop}%
\bibitem [{\citenamefont {{van Veen}}\ and\ \citenamefont
  {Kawahara}(2011)}]{vanveenHomoclinicTangleEdge2011}%
  \BibitemOpen
  \bibfield  {author} {\bibinfo {author} {\bibfnamefont {L.}~\bibnamefont {{van
  Veen}}}\ and\ \bibinfo {author} {\bibfnamefont {G.}~\bibnamefont
  {Kawahara}},\ }\bibfield  {title} {\bibinfo {title} {Homoclinic {{Tangle}} on
  the {{Edge}} of {{Shear Turbulence}}},\ }\href@noop {} {\bibfield  {journal}
  {\bibinfo  {journal} {Phys. Rev. Lett.}\ }\textbf {\bibinfo {volume} {107}},\
  \bibinfo {pages} {114501} (\bibinfo {year} {2011})}\BibitemShut {NoStop}%
\bibitem [{\citenamefont {Doering}\ and\ \citenamefont
  {Constantin}(1992)}]{doeringEnergyDissipationShear1992}%
  \BibitemOpen
  \bibfield  {author} {\bibinfo {author} {\bibfnamefont {C.~R.}\ \bibnamefont
  {Doering}}\ and\ \bibinfo {author} {\bibfnamefont {P.}~\bibnamefont
  {Constantin}},\ }\bibfield  {title} {\bibinfo {title} {Energy dissipation in
  shear driven turbulence},\ }\href@noop {} {\bibfield  {journal} {\bibinfo
  {journal} {Phys. Rev. Lett.}\ }\textbf {\bibinfo {volume} {69}},\ \bibinfo
  {pages} {1648} (\bibinfo {year} {1992})}\BibitemShut {NoStop}%
\bibitem [{\citenamefont {Waleffe}(2011)}]{waleffeOverviewTurbulentShear2011}%
  \BibitemOpen
  \bibfield  {author} {\bibinfo {author} {\bibfnamefont {F.}~\bibnamefont
  {Waleffe}},\ }\bibfield  {title} {\bibinfo {title} {Overview of turbulent
  shear flows},\ }in\ \href@noop {} {\emph {\bibinfo {booktitle} {Summer
  {{Program}} in {{Geophysical Fluid Dynamics}}}}}\ (\bibinfo  {publisher}
  {{Woods Hole Oceanographic Institution}},\ \bibinfo {year}
  {2011})\BibitemShut {NoStop}%
\bibitem [{\citenamefont {Wang}\ \emph {et~al.}(2007)\citenamefont {Wang},
  \citenamefont {Gibson},\ and\ \citenamefont
  {Waleffe}}]{wangLowerBranchCoherent2007}%
  \BibitemOpen
  \bibfield  {author} {\bibinfo {author} {\bibfnamefont {J.}~\bibnamefont
  {Wang}}, \bibinfo {author} {\bibfnamefont {J.}~\bibnamefont {Gibson}},\ and\
  \bibinfo {author} {\bibfnamefont {F.}~\bibnamefont {Waleffe}},\ }\bibfield
  {title} {\bibinfo {title} {Lower {{Branch Coherent States}} in {{Shear
  Flows}}: {{Transition}} and {{Control}}},\ }\href@noop {} {\bibfield
  {journal} {\bibinfo  {journal} {Phys. Rev. Lett.}\ }\textbf {\bibinfo
  {volume} {98}},\ \bibinfo {pages} {204501} (\bibinfo {year}
  {2007})}\BibitemShut {NoStop}%
\bibitem [{\citenamefont {Skufca}\ \emph {et~al.}(2006)\citenamefont {Skufca},
  \citenamefont {Yorke},\ and\ \citenamefont
  {Eckhardt}}]{skufcaEdgeChaosParallel2006}%
  \BibitemOpen
  \bibfield  {author} {\bibinfo {author} {\bibfnamefont {J.~D.}\ \bibnamefont
  {Skufca}}, \bibinfo {author} {\bibfnamefont {J.~A.}\ \bibnamefont {Yorke}},\
  and\ \bibinfo {author} {\bibfnamefont {B.}~\bibnamefont {Eckhardt}},\
  }\bibfield  {title} {\bibinfo {title} {Edge of {{Chaos}} in a {{Parallel
  Shear Flow}}},\ }\href@noop {} {\bibfield  {journal} {\bibinfo  {journal}
  {Phys. Rev. Lett.}\ }\textbf {\bibinfo {volume} {96}},\ \bibinfo {pages}
  {174101} (\bibinfo {year} {2006})}\BibitemShut {NoStop}%
\bibitem [{\citenamefont {Schneider}\ \emph {et~al.}(2008)\citenamefont
  {Schneider}, \citenamefont {Gibson}, \citenamefont {Lagha}, \citenamefont
  {De~Lillo},\ and\ \citenamefont
  {Eckhardt}}]{schneiderLaminarturbulentBoundaryPlane2008}%
  \BibitemOpen
  \bibfield  {author} {\bibinfo {author} {\bibfnamefont {T.~M.}\ \bibnamefont
  {Schneider}}, \bibinfo {author} {\bibfnamefont {J.~F.}\ \bibnamefont
  {Gibson}}, \bibinfo {author} {\bibfnamefont {M.}~\bibnamefont {Lagha}},
  \bibinfo {author} {\bibfnamefont {F.}~\bibnamefont {De~Lillo}},\ and\
  \bibinfo {author} {\bibfnamefont {B.}~\bibnamefont {Eckhardt}},\ }\bibfield
  {title} {\bibinfo {title} {Laminar-turbulent boundary in plane {{Couette}}
  flow},\ }\href@noop {} {\bibfield  {journal} {\bibinfo  {journal} {Phys. Rev.
  E}\ }\textbf {\bibinfo {volume} {78}} (\bibinfo {year} {2008})}\BibitemShut
  {NoStop}%
\bibitem [{\citenamefont {Avila}\ \emph {et~al.}(2013)\citenamefont {Avila},
  \citenamefont {Mellibovsky}, \citenamefont {Roland},\ and\ \citenamefont
  {Hof}}]{avilaStreamwiseLocalizedSolutionsOnset2013}%
  \BibitemOpen
  \bibfield  {author} {\bibinfo {author} {\bibfnamefont {M.}~\bibnamefont
  {Avila}}, \bibinfo {author} {\bibfnamefont {F.}~\bibnamefont {Mellibovsky}},
  \bibinfo {author} {\bibfnamefont {N.}~\bibnamefont {Roland}},\ and\ \bibinfo
  {author} {\bibfnamefont {B.}~\bibnamefont {Hof}},\ }\bibfield  {title}
  {\bibinfo {title} {Streamwise-{{Localized Solutions}} at the {{Onset}} of
  {{Turbulence}} in {{Pipe Flow}}},\ }\href@noop {} {\bibfield  {journal}
  {\bibinfo  {journal} {Phys. Rev. Lett.}\ }\textbf {\bibinfo {volume} {110}},\
  \bibinfo {pages} {224502} (\bibinfo {year} {2013})}\BibitemShut {NoStop}%
\bibitem [{\citenamefont {Halcrow}\ \emph {et~al.}(2009)\citenamefont
  {Halcrow}, \citenamefont {Gibson}, \citenamefont {Cvitanovi{\'c}},\ and\
  \citenamefont {Viswanath}}]{halcrowHeteroclinicConnectionsPlane2009}%
  \BibitemOpen
  \bibfield  {author} {\bibinfo {author} {\bibfnamefont {J.}~\bibnamefont
  {Halcrow}}, \bibinfo {author} {\bibfnamefont {J.~F.}\ \bibnamefont {Gibson}},
  \bibinfo {author} {\bibfnamefont {P.}~\bibnamefont {Cvitanovi{\'c}}},\ and\
  \bibinfo {author} {\bibfnamefont {D.}~\bibnamefont {Viswanath}},\ }\bibfield
  {title} {\bibinfo {title} {Heteroclinic connections in plane {{Couette}}
  flow},\ }\href@noop {} {\bibfield  {journal} {\bibinfo  {journal} {J. Fluid
  Mech.}\ }\textbf {\bibinfo {volume} {621}},\ \bibinfo {pages} {365} (\bibinfo
  {year} {2009})}\BibitemShut {NoStop}%
\bibitem [{\citenamefont
  {Viswanath}(2007)}]{viswanathRecurrentMotionsPlane2007}%
  \BibitemOpen
  \bibfield  {author} {\bibinfo {author} {\bibfnamefont {D.}~\bibnamefont
  {Viswanath}},\ }\bibfield  {title} {\bibinfo {title} {Recurrent motions
  within plane {{Couette}} turbulence},\ }\href@noop {} {\bibfield  {journal}
  {\bibinfo  {journal} {J. Fluid Mech.}\ }\textbf {\bibinfo {volume} {580}},\
  \bibinfo {pages} {339} (\bibinfo {year} {2007})}\BibitemShut {NoStop}%
  \bibitem [{\citenamefont {Whitney}(1944)}]{whitney}%
  \BibitemOpen
  \bibfield  {author} {\bibinfo {author} {\bibfnamefont {H.}~\bibnamefont
  {Whitney}},\ }\bibfield  {title} { {\bibinfo {title} {The
  self-intersections of a smooth n-manifold in 2n-space},\ }}\href@noop {}
  {\bibfield  {journal} {\bibinfo  {journal} {Annals of Mathematics}\ }\textbf
  {\bibinfo {volume} {45}},\ \bibinfo {pages} {220} (\bibinfo {year}
  {1944})}\BibitemShut {NoStop}%
\bibitem [{\citenamefont {Dankowicz}\ and\ \citenamefont
  {Schilder}(2013)}]{coco}%
  \BibitemOpen
  \bibfield  {author} {\bibinfo {author} {\bibfnamefont {H.}~\bibnamefont
  {Dankowicz}}\ and\ \bibinfo {author} {\bibfnamefont {F.}~\bibnamefont
  {Schilder}},\ }\href@noop {} {\emph {\bibinfo {title} {Recipes for
  Continuation}}}\ (\bibinfo  {publisher} {SIAM},\ \bibinfo {address} {USA},\ \bibinfo {year} {2013})\BibitemShut
  {NoStop}%
  \bibitem [{\citenamefont {Guckenheimer}\ and\ \citenamefont
  {Holmes}(1983)}]{guckenheimerNonlinearOscillationsDynamical1983}%
  \BibitemOpen
    \bibfield  {author} {\bibinfo {author} {\bibfnamefont {J.}~\bibnamefont
  {Guckenheimer}}\ and\ \bibinfo {author} {\bibfnamefont {P.}~\bibnamefont
  {Holmes}},\ }\href@noop {} {\emph {\bibinfo {title} {Nonlinear
  {{Oscillations}}, {{Dynamical Systems}}, and {{Bifurcations}} of {{Vector
  Fields}}}}},\ \bibinfo {series} {Applied {{Mathematical Sciences}}},
  Vol.~\bibinfo {volume} {42}\ (\bibinfo  {publisher} {{Springer}},\
  \bibinfo {address} {{Berlin}},\ \bibinfo {year} {2013})\BibitemShut
  {NoStop}%
\bibitem [{\citenamefont
  {Stuart}(1958)}]{stuartNonlinearMechanicsHydrodynamic1958}%
  \BibitemOpen
  \bibfield  {author} {\bibinfo {author} {\bibfnamefont {J.~T.}\ \bibnamefont
  {Stuart}},\ }\bibfield  {title} {\bibinfo {title} {On the non-linear
  mechanics of hydrodynamic stability},\ }\href@noop {} {\bibfield  {journal}
  {\bibinfo  {journal} {J. Fluid Mech.}\ }\textbf {\bibinfo {volume} {4}},\
  \bibinfo {pages} {1} (\bibinfo {year} {1958})}\BibitemShut {NoStop}%
\bibitem [{\citenamefont {Landau}(1959)}]{landauFluidMechanics1959}%
  \BibitemOpen
  \bibfield  {author} {\bibinfo {author} {\bibfnamefont {L.~D.}\ \bibnamefont
  {Landau}} and \bibinfo {author} {\bibfnamefont {E.~M.}\ \bibnamefont
  {Lifshitz}},\ }\href@noop {} {\emph {\bibinfo {title} {Fluid Mechanics}}}\
  (\bibinfo  {publisher} {{Pergamon Press, Oxford}},\ \bibinfo {year} {1987})\BibitemShut
  {NoStop}%
\bibitem [{\citenamefont {Ashwin}\ \emph {et~al.}(2012)\citenamefont {Ashwin},
  \citenamefont {Wieczorek}, \citenamefont {Vitolo},\ and\ \citenamefont
  {Cox}}]{ashwinTippingPointsOpen2012a}%
  \BibitemOpen
  \bibfield  {author} {\bibinfo {author} {\bibfnamefont {P.}~\bibnamefont
  {Ashwin}}, \bibinfo {author} {\bibfnamefont {S.}~\bibnamefont {Wieczorek}},
  \bibinfo {author} {\bibfnamefont {R.}~\bibnamefont {Vitolo}},\ and\ \bibinfo
  {author} {\bibfnamefont {P.}~\bibnamefont {Cox}},\ }\bibfield  {title}
  {\bibinfo {title} {Tipping points in open systems: Bifurcation, noise-induced
  and rate-dependent examples in the climate system},\ }\href@noop {}
  {\bibfield  {journal} {\bibinfo  {journal} {Phil. Trans. R. Soc. A.}\
  }\textbf {\bibinfo {volume} {370}},\ \bibinfo {pages} {1166} (\bibinfo {year}
  {2012})}\BibitemShut {NoStop}%
\bibitem [{\citenamefont {Alkhayuon}\ and\ \citenamefont
  {Ashwin}(2018)}]{alkhayuonRateinducedTippingPeriodic2018}%
  \BibitemOpen
  \bibfield  {author} {\bibinfo {author} {\bibfnamefont {H.~M.}\ \bibnamefont
  {Alkhayuon}}\ and\ \bibinfo {author} {\bibfnamefont {P.}~\bibnamefont
  {Ashwin}},\ }\bibfield  {title} {\bibinfo {title} {Rate-induced tipping from
  periodic attractors: {{Partial}} tipping and connecting orbits},\ }\href@noop
  {} {\bibfield  {journal} {\bibinfo  {journal} {Chaos}\ }\textbf {\bibinfo
  {volume} {28}},\ \bibinfo {pages} {033608} (\bibinfo {year}
  {2018})}\BibitemShut {NoStop}%
\bibitem [{\citenamefont {Kreilos}\ and\ \citenamefont
  {Eckhardt}(2012)}]{kreilosPeriodicOrbitsOnset2012}%
  \BibitemOpen
  \bibfield  {author} {\bibinfo {author} {\bibfnamefont {T.}~\bibnamefont
  {Kreilos}}\ and\ \bibinfo {author} {\bibfnamefont {B.}~\bibnamefont
  {Eckhardt}},\ }\bibfield  {title} {\bibinfo {title} {Periodic orbits near
  onset of chaos in plane {{Couette}} flow},\ }\href@noop {} {\bibfield
  {journal} {\bibinfo  {journal} {Chaos}\ }\textbf {\bibinfo {volume} {22}},\
  \bibinfo {pages} {047505} (\bibinfo {year} {2012})}\BibitemShut {NoStop}%
\end{thebibliography}
\end{document}



\title{Supplemental Material: \\ Dynamics-based machine learning of transitions in Couette flow}
\author{Bálint Kaszás, Mattia Cenedese, George Haller \\ Institute for Mechanical Systems, ETH Zürich, Zürich, Switzerland}
\date{}

\maketitle
\section{Data availability}
All trajectories we use in this Letter are generated with the open-source computational fluid dynamics library \texttt{Channelflow} \cite{channelflow}. The velocity field $\mathbf{u}(\mathbf{x},t)$ is stored in discrete spatial locations $\mathbf{u}(x_j, y_j, z_j,t)$ due to the spectral discretization. A sample of the generated data is available under the link \cite{ssmlearnCouette}. In addition to the raw data, we also made available the codes used for the analysis in the form of a commented MATLAB\textsuperscript{\tiny\textregistered} live script, as a part of the open-source toolbox \texttt{SSMLearn} \cite{cenedeseDatadrivenModelingPrediction2022}. The full data set is available from the authors upon request.

\section{Spectral Submanifolds}

In this section, we briefly recall the properties of spectral submanifolds (SSMs) in general dynamical systems. 

Based on the two periodic directions (along $x$ and $z$), we discretize the incompressible Navier--Stokes equations  \cite{channelflow} by Galerkin-projection onto a Fourier-Chebyshev basis, 

\begin{equation}
\label{spectralexpansion}
\mathbf{u}(x,y,z,t) = \sum_{k_x, k_z, n_y} \hat{\mathbf{u}}_{k_x,k_z,n_y}(t)T_{n_y}(y)e^{2\pi i\left(k_x x/Lx + k_z z / L_z \right)},
\end{equation}

where $T_{m}(y)$ is the $m$-th Chebyshev polynomial. The sums in \eqref{spectralexpansion} are taken over a finite number of wave-numbers denoted by $k_z, k_x$ and Chebyshev-modes indexed by $n_y$. With this discretization, the time-evolution of the spectral coefficients is governed by the following ordinary differential equation

\begin{equation}
\label{discreteNS}
    \frac{d}{dt}\hat{\mathbf{u}} = \mathbf{A}\hat{\mathbf{u}} + \mathbf{f}_{nl}(\hat{\mathbf{u}}) \;\; \mathbf{f}(\hat{\mathbf{u}}) \in O(|\hat{\mathbf{u}}|^2),
\end{equation}

where we have introduced the single vector $\hat{\mathbf{u}}\in \mathbb{R}^N$ to be collection of all spectral coefficients of all three velocity components. The matrix $\mathbf{A}\in \mathbb{R}^{N\times N}$ represents the linear part of the dynamics, and all nonlinear terms are collected in $\mathbf{f}(\hat{\mathbf{u}})$, which is a $C^\infty$ smooth function. The matrix $\mathbf{A}$ and the coefficients of $\mathbf{f}$ can be obtained by substituting \eqref{spectralexpansion} in the Navier--Stokes equations and enforcing incompressibility to eliminate the pressure $p$.

Equation \eqref{discreteNS} describes a finite-dimensional dynamical system, to which the results of \cite{hallerNonlinearNormalModes2016a} are applicable. Those results do extend to infinite dimensions \cite{cabreParameterizationMethodInvariant2003a} but require technical assumptions that would be challenging to verify for the Navier-Stokes equations \cite{kogelbauerRigorousModelReduction2018}.

Let us assume that $\hat{\mathbf{u}}_0$ is a hyperbolic fixed point of \eqref{discreteNS}. In our case, this could represent the laminar base state, the lower- or upper branch Nagata invariant solutions. Because of the hyperbolicity of the fixed points representing these states in the phase space, the local dynamics around $\hat{\mathbf{u}}$  is determined by the linear part, which is the matrix $\mathbf{A}$. We define a  spectral subspace as the vector space spanned by a set of eigenvectors of $\mathbf{A}$. Any such subspace is invariant under the linearized dynamics.  Important special cases of spectral subspaces are the stable and unstable subspace, which are spanned by the eigenvectors of $\mathbf{A}$, whose eigenvalues have strictly negative and strictly positive real parts, respectively. By the center manifold theorem \cite{guckenheimerNonlinearOscillationsDynamical1983}, the nonlinear system \eqref{discreteNS} has corresponding invariant manifolds (called stable and unstable manifolds) that are tangent to the unstable and stable subspaces at the fixed point. 

As discussed in \cite{cabreParameterizationMethodInvariant2003a, hallerNonlinearNormalModes2016a}, such stable and unstable manifolds are further foliated by lower-dimensional invariant submanifolds that are tangent to spectral subspaces of the unstable and stable subspaces. We focus here on the slowest spectral subspaces spanned by the eigenvectors associated to the eigenvalues with decay rates closest to zero. We denote by $E_d$ such a $d$-dimensional spectral subspace, where $d=1, 2$ will hold in the present paper.
A spectral submanifold (SSM) will then be a  $d-$dimensional invariant manifold, $W_d$ of the full nonlinear system that is tangent to $E_d$ at the fixed point $\hat{\mathbf{u}}_0$. There is an abundance of such invariant manifolds already in the linearized system, so the non-uniqueness of $W_d$ in the nonlinear system is fully expected. Nevertheless, building on prior work by \cite{cabreParameterizationMethodInvariant2003a}, the theorems of \cite{hallerNonlinearNormalModes2016a} establish the existence of a unique unique smoothest $W_d$ under appropriate non-resonance conditions on the spectrum $\mathrm{Spect }\,\mathbf{A}$ of $\mathbf{A}$.  

To state these results, we first define the spectral quotient corresponding to the spectral subspaces $E_d$   as 

\begin{equation}
    \sigma(E_d) = \text{Int }\left[\frac{\max_{\lambda \in \mathrm{Spect }\,\mathbf{A}} |\text{Re }\lambda |}{\min_{\lambda \in \mathrm{Spect }\,\mathbf{A}_{E_d}} |\text{Re }\lambda |}\right],
\end{equation}
where $\mathrm{Spect }\,\mathbf{A}_{E_d}= \{ \lambda_1, ..., \lambda_d \}$ denotes the  restriction of the spectrum of $\mathbf{A}$ to the spectral subspace $E_d$. In our present setting,  \eqref{discreteNS} arises from the discretization of the Navier--Stokes equations, and hence its linear part has eigenvalues with arbitrarily large negative real parts \cite{orrStabilityInstabilitySteady1907}, making $\sigma(E_d)$   infinite.  In practice, $\sigma(E_d)$ is a large positive number of the order of a $100$, whose exact value  depends on the level of the discretization used in obtaining eq. \eqref{discreteNS}.  
 
We call the spectral subspace $E_d$  non-resonant if 
\begin{equation}
        \label{nonresonance}
    \sum_{i=1}^d m_i \lambda_i \neq \lambda_j, \quad \lambda_j \in \mathrm{Spect }\,\mathbf{A} - \{ \lambda_1, ..., \lambda_d \}
\end{equation}
hold for any set of nonnegative integers  $m_i \in \mathbb{N}$ satisfying $2 \leq \sum_{i=1}^d m_i \leq \sigma(E_d)$. We note that the non-resonance conditions stated here are generically satisfied, i.e., hold with probability 1. Even if the spectrum of the underlying PDE were resonant due to symmetries, these would not be exact resonances in the finite-dimensional truncation \eqref{discreteNS}.

The main result of SSM theory is that for any nonresonant $E$, a (primary)  SSM defined above exists that is unique among all  $C^{\sigma(E_d) +1 }$ invariant manifolds tangent to $E_d$ at the fixed point $\hat{\mathbf{u}}_0$. All other invariant manifolds (secondary SSMs) tangent to $E_d$ at $\hat{\mathbf{u}}_0$ are only $C^r$ smooth with $r < \sigma(E_d)+1$. 

We note that although the finite-dimensional representation of the Navier--Stokes was formulated in terms of the spectral coefficients, the SSMs can also be found in the phase space of velocity measurements. By Whitney’s embedding theorem \cite{whitney}, the SSMs will be smoothly embedded in the high-dimensional space of the velocity measurements with probability $1$. 

\section{Parametrizing spectral submanifolds with the rate of energy input and the dissipation}
From simulations we note a one-to-one correspondence between the dynamics on the SSMs and that of $I$ and $(I,D)$, for the one- and two-dimensional SSMs respectively. Hence, we can use these variables to parametrize the SSMs for the full velocity field. In the main text, we choose $J = \sqrt{|I|}$ and $K =\sqrt{|D|}$ instead of $I$ and $D$.

\begin{figure}[b!]
    \centering
    \includegraphics[width=\textwidth]{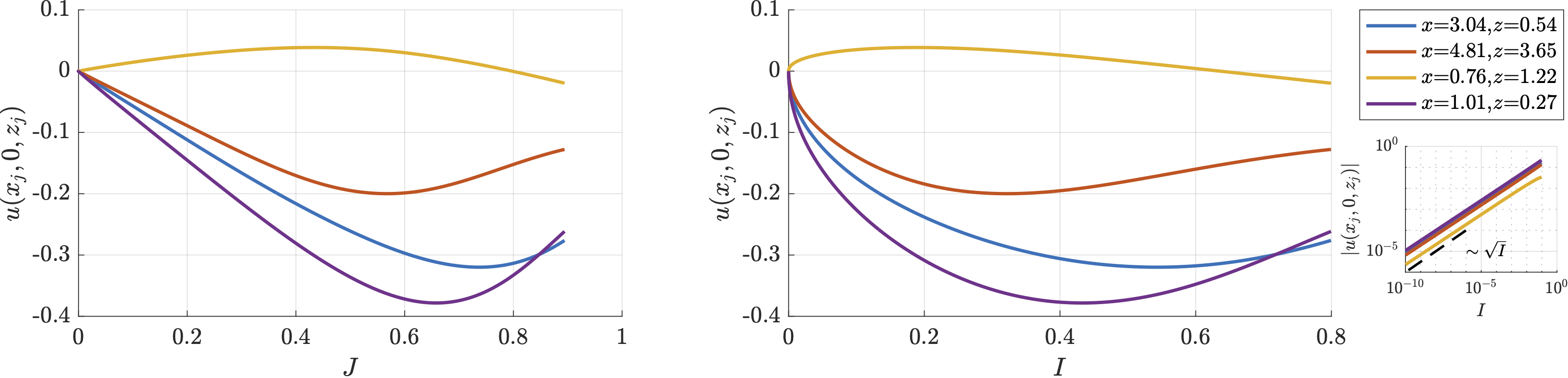}
    \caption{Dependence of the streamwise velocity $u$ on the energy input $I$ and on its square root $J$, measured at 4 points in the channel. The trajectory is initialized on the unstable manifold of the lower-branch and converges to the base state, at $\text{Re} = 134.52$. The inset in the lower right shows the absolute values of the velocity measurements as a function of $I$ on logarithmic scales. }
    \label{sfig1}
\end{figure}

To motivate our choice, we consider the case in which the SSM is one-dimensional, but similar considerations hold for the other regimes investigated in this Letter. Figure \ref{sfig1} shows the dependence of the flow field on the coordinates $I$ and $J$. We display the streamwise velocity (the $u$ component) in 4 spatial locations in the channel. The inset in the lower right of Fig. \ref{sfig1} shows a power-law type behavior of $|u(x_j, 0, z_j)|(I)$, with exponent $1/2$.

Note that in the left panel of Fig. \ref{sfig1}, the trajectory has a vertical tangent at the origin, and hence the velocity components near $I=0$ cannot be expressed as a smooth graph over $I$. Due to this singularity, the derivative with respect to $I$ diverges. The right panel shows that viewing the flow field as a function over $J$ resolves this issue as the graph no longer has a vertical tangent at $I=J=0$. Therefore, the vector-valued coefficients $\mathbf{w}$ in the parametrizations of the SSM can be expressed as
\begin{align*}
        \mathbf{u}(x,y,z,t) &= \mathbf{u}_\text{base} + \sum_{l=1}^{M_p}\sum_{m=0}^1  \mathbf{w}_{l,m}(x,y,z) J^l(t)( \text{Re} - \text{Re}_c)^m \quad \text{in region (I),}\label{eq:param1}    \\ 
    \mathbf{u}(x,y,z,t) &=\mathbf{u}_\text{base} + \sum_{\substack{l,m \\ l+m \leq M_p}} \mathbf{w}_{l,m}(x,y,z) J^l(t) K^m(t) \quad \text{in regions (II) and (III)},
\end{align*}
where we have parametric dependence in Re, measured from a reference value Re$_c$, for regime (I). 

To identify the coefficients of the parametrization, we minimize the $L^2$ norm of the difference between the velocity field $\mathbf{u}$ and its reconstruction. That is, the coefficients are obtained by minimizing the error

\begin{align*}
    \mathbf{w}^*_{l,m} &= \text{arg}\min_{\mathbf{w}_{l,m}} \sum_{j=1}^{N_{dp}}\left \Vert \mathbf{u}(x,y,z, t_j) - \mathbf{u}_{\text{base}} -  \sum_{q=1}^{M_p}\sum_{r=0}^1  \mathbf{w}_{q,r}(x,y,z) J^q(t_j)( \text{Re} - \text{Re}_c)^r\right \Vert^2_{L^2}, \;\; \text{or } \\
    \mathbf{w}^*_{l,m} &= \text{arg}\min_{\mathbf{w}_{l,m}} \sum_{j=1}^{N_{dp}}\left \Vert \mathbf{u}(x,y,z, t_j) - \mathbf{u}_{\text{base}} -  \sum_{\substack{q,r \\ q+r \leq M_p}}^{M_p} \mathbf{w}_{q,r}(x,y,z) J^q(t_j)K^r(t_j)\right \Vert^2_{L^2},
\end{align*}

for the $N_{dp}$ data-snapshots available, whose solution is that of a least-squares minimization problem \cite{hastie2009elements}. 

\section{Additional details of the reduced dynamical models \label{appc}}

We use reduced dynamics of the form 
\begin{align*}
        J_{n+1} &= R(J_{n}, \text{Re}) \quad \text{in region (I)}, \\
    \begin{pmatrix} J_{n+1} \\ K_{n+1}\end{pmatrix}  &= \begin{pmatrix} R_1(J_n, K_n) \\ R_2(J_n, K_n) \end{pmatrix} \quad \text{in region (II).}
\end{align*}
In region (I), we allow for Reynolds-number-dependent coefficients in the polynomial expansion of the function $R$ up to first order. We look for $R$ in the form
\begin{equation}\label{eq:rdynamicsI}
    R(J, \text{Re}) = \sum_{l=1}^{M_d}\sum_{m=0}^{1} R_{l,m} J^l (\text{Re}- \text{Re}_c)^m,
\end{equation}
where $\text{Re}_c=134.5$ is a reference value for the Reynolds number, around which we take the expansion. 
In the two-dimensional regime (II), the reduced dynamics is 
\begin{equation}\label{eq:rdynamicsII}
    R_{1,2}(J,K) = \sum_{\substack{l,m \\ l+m\leq M_d}} R^{(1,2)}_{l,m} J^m K^l.
\end{equation}

In all cases, we denoted the maximal order in the polynomial expansions by $M_d$. The coefficients of the reduced dynamics are obtained from data by \texttt{SSMLearn}, and, for cases (I) and (II), the a priori knowledge of non-trivial fixed points is given to the model identification. Specifically, the discrete time dynamical system $\boldsymbol{\eta}_{n+1} = \mathbf{r}(\boldsymbol{\eta}_{n},\boldsymbol{\mu})$ on the state variable $\boldsymbol{\eta}\in\mathbb{R}^m$ depending on the parameters $\boldsymbol{\mu}\in\mathbb{R}^d$ and with $N_{fp}$ fixed points can be identified from $N_{dp}$ data points by solving the constrained minimization problem
\begin{equation}\label{eq:constrainedregression}
\mathbf{r}^* = \arg\min \sum_{j=1}^{N_{dp}-1}\| \boldsymbol{\eta}_{j+1} - \mathbf{r}(\boldsymbol{\eta}_j,\boldsymbol{\mu}_j) \|^2 + \sum_{l=1}^{N_{fp}}  2\boldsymbol{\lambda}_l^\top (\mathbf{r}(\boldsymbol{\eta}_{fp,l},\boldsymbol{\mu}_{fp,l}) - \boldsymbol{\eta}_{fp,l}) + C_p(\mathbf{r}),
\end{equation}
where $\boldsymbol{\lambda}_l$ for $l=1,2,...N_{fp}$ are the Lagrange multipliers and $C_p(\mathbf{r})$ is a regularization term. For the polynomial representations of $\mathbf{r}$ in Eqs. (\ref{eq:rdynamicsI},\ref{eq:rdynamicsII}) and a ridge-type penalty, their coefficients are identified by solving problem (\ref{eq:constrainedregression}) in closed form. Specifically, this representation takes the form $\mathbf{r}(\boldsymbol{\eta},\boldsymbol{\mu}) = \boldsymbol{\eta} + \mathbf{R}\boldsymbol{\varphi}(\boldsymbol{\eta},\boldsymbol{\mu})$ where the feature map $\boldsymbol{\varphi}$ denotes the selected $N_{m}$ multivariate monomials in $(\boldsymbol{\eta},\boldsymbol{\mu})$ and $\mathbf{R}\in\mathbb{R}^{m\times N_{m}}$ is a matrix containing their coefficients. As we center around the base (laminar) state, we need to assume that $\mathbf{0} = \mathbf{r}(\mathbf{0},\boldsymbol{\mu})$, which implies that the monomials have at least linear dependence on the coordinates $\boldsymbol{\eta}$. By indicating the $k$-th component of a vector $\boldsymbol{\eta}$ as $\boldsymbol{\eta}^{(k)}$, we define 
\begin{equation}
    \begin{array}{l}
    \mathbf{R}^\top = \begin{bmatrix} \mathbf{r}_1 & \mathbf{r}_2 & ... & \mathbf{r}_m \end{bmatrix}, \qquad 
    \mathbf{X}^\top = \begin{bmatrix} \boldsymbol{\varphi}(\boldsymbol{\eta}_1,\boldsymbol{\mu}_1) &  \boldsymbol{\varphi}(\boldsymbol{\eta}_2,\boldsymbol{\mu}_2) & ... & \boldsymbol{\varphi}(\boldsymbol{\eta}_{N_{dp}-1},\boldsymbol{\mu}_{N_{dp}-1}) \end{bmatrix}, \\
    \mathbf{y}_k^\top = \begin{pmatrix} \boldsymbol{\eta}^{(k)}_2-\boldsymbol{\eta}^{(k)}_1 &  \boldsymbol{\eta}^{(k)}_3-\boldsymbol{\eta}^{(k)}_2 & ... & \boldsymbol{\eta}^{(k)}_{N_{dp}}-\boldsymbol{\eta}^{(k)}_{N_{dp}-1} \end{pmatrix}, \,\,  \mathbf{Y} =\begin{bmatrix} \mathbf{y}_1 & \mathbf{y}_2 & ... & \mathbf{y}_{m} \end{bmatrix}, \\
    \mathbf{X}_{fp}^\top = \begin{bmatrix} \boldsymbol{\varphi}(\boldsymbol{\eta}_{fp,1},\boldsymbol{\mu}_{fp,1}) &  \boldsymbol{\varphi}(\boldsymbol{\eta}_{fp,2},\boldsymbol{\mu}_{fp,2}) & ... & \boldsymbol{\varphi}(\boldsymbol{\eta}_{fp,N_{fp}},\boldsymbol{\mu}_{fp,N_{fp}}) \end{bmatrix}, \\
    \hat{\boldsymbol{\lambda}}^\top_k = \begin{pmatrix} \boldsymbol{\lambda}_1^{(k)} & \boldsymbol{\lambda}_2^{(k)} & ... & \boldsymbol{\lambda}_{N_{dp}-1}^{(k)} \end{pmatrix}, \qquad \boldsymbol{\Lambda} = \begin{bmatrix} \hat{\boldsymbol{\lambda}}_1 & \hat{\boldsymbol{\lambda}}_2 & ... & \hat{\boldsymbol{\lambda}}_m \end{bmatrix},
    \end{array}
\end{equation}
and let $\mathbf{L}\in\mathbb{R}^{N_m\times N_m}$ a diagonal matrix whose values are the maximum absolute values taken along the columns of $\mathbf{X}\in\mathbb{R}^{N_{dp}\times N_m}$. Then, we can reformulate problem (\ref{eq:constrainedregression}) using the cost function
\begin{equation}\label{eq:constrainedregression2}
    C(\mathbf{R},\boldsymbol{\Lambda};\lambda)= \sum_{k=1}^{m} \left\| \mathbf{y}_k - \mathbf{X}\mathbf{r}_k \right\|^2 + \langle \hat{\boldsymbol{\lambda}}_k, \mathbf{X}_{fp}\mathbf{r}_k\rangle + \lambda \| \mathbf{L}\mathbf{r}_k\|^2,
\end{equation}
as $(\mathbf{R}^*,\boldsymbol{\Lambda}^*)=\arg\min C(\mathbf{R},\boldsymbol{\Lambda};\lambda)$, whose solution is
\begin{equation}
    \begin{pmatrix} \mathbf{R}^{*\top} \\ \boldsymbol{\Lambda}^* \end{pmatrix} = \begin{bmatrix}\mathbf{X}^\top\mathbf{X} + \lambda\mathbf{L}^2 & \mathbf{X}_{fp}^\top \\ \mathbf{X}_{fp} & \mathbf{0} \end{bmatrix}^{-1} \begin{pmatrix}\mathbf{X}^\top\mathbf{Y} \\ \mathbf{0} \end{pmatrix},
\end{equation}
where the parameter $\lambda$ is determined via cross-validation \cite{hastie2009elements}. We remark that analogous developments hold for the case of a vector field $\dot{\boldsymbol{\eta}} = \mathbf{r}(\boldsymbol{\eta},\boldsymbol{\mu}) = \mathbf{R}\boldsymbol{\varphi}(\boldsymbol{\eta},\boldsymbol{\mu})$, in which one would only need to set $\mathbf{y}_k$ in (\ref{eq:constrainedregression2}) as a vector of time derivatives. Table \ref{tab1} summarizes the maximal polynomial orders for the parametrization $M_p$ and for the reduced dynamics $M_d$ with which the results in the main text were obtained.
\begin{table}\label{tab1}
\begin{center}
\begin{tabular}{ |c|c|c|c| } 
\hline
& Region (I) & Region (II) & Region (III) \\
 \hline
$M_d$ & 15 & 4 & 5 \\ 
$M_p$ & 6 & 2 & 5 \\ 
 \hline
\end{tabular}
\end{center}
\caption{Maximal polynomial orders used in the expressions of the reduced dynamics ($M_d$) and in the parametrization of the SSM ($M_p$). }
\end{table}
\begin{figure}[b!]
    \centering
    \includegraphics[width=\textwidth]{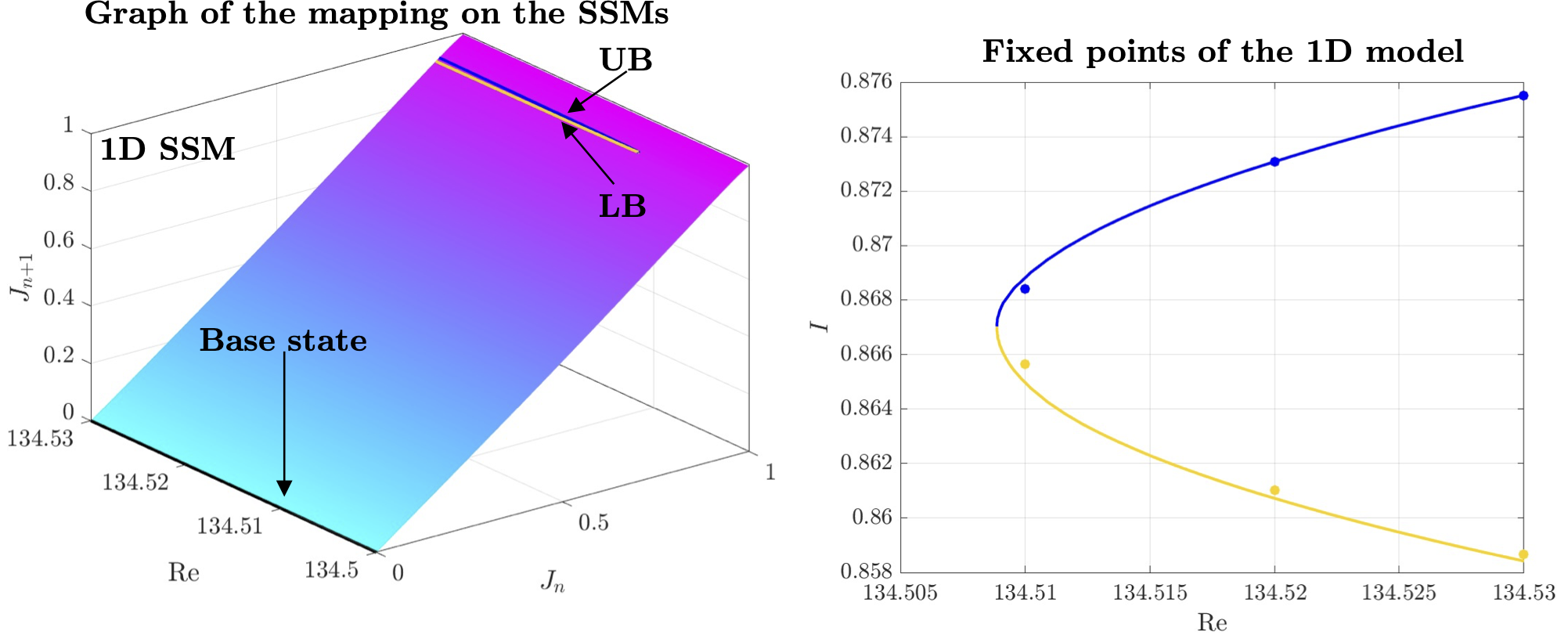}
    \caption{Left panel: the mapping $J_{n+1} = R(J_n, \text{Re})$ on the one-dimensional SSM with the three curves of fixed points are shown. Right panel: lower- and upper-branch fixed points in the ($I,\text{Re}$) plane obtained by numerical continuation. The blue curve shows the stable branch and the unstable branch is plotted in yellow. Colored circles indicate the Reynolds numbers at which training trajectories were initialized. }
    \label{sfig2}
\end{figure}
\begin{figure}[b!]
    \centering
    \includegraphics[width=0.5\textwidth]{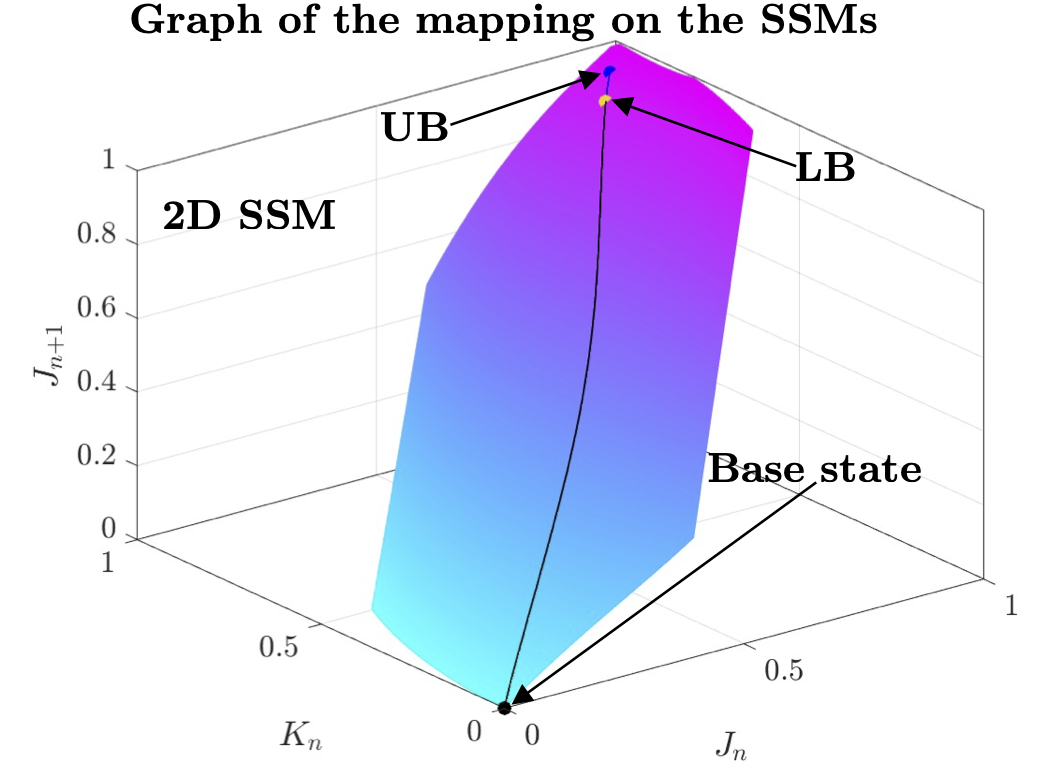}
    \caption{The mapping $J_{n+1} = R_1(J_n, K_n)$ on the two-dimensional SSM. Colored circles indicate the three fixed points. }
    \label{sfig3}
\end{figure}
\begin{figure}[t!]
    \centering
    \includegraphics[width=\textwidth]{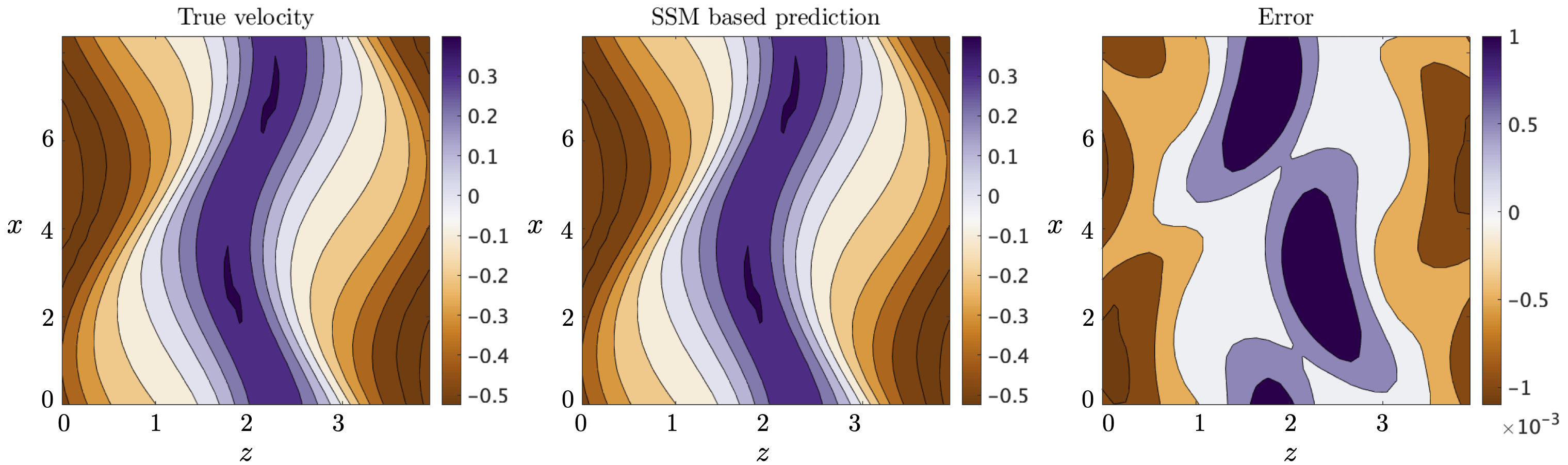}
    \caption{Left panel: streamwise component in the $y=0$ plane of a velocity field at $t=184$ which was initialized off the SSM at $\text{Re}=134.52$. Middle panel: SSM-based prediction of the same velocity field, obtained as a composition of the reduced dynamics and the parametrization. Right panel: difference between the true velocity field and the SSM-based prediction. }
    \label{sfig4}
\end{figure}

Figure \ref{sfig2}. shows the mapping $R(J,\text{Re})$. In the left panel, we show its value over the $(J, \text{Re})$ plane. For training, we use a total of 6 trajectories, which are initialized on the unstable manifold of the lower-branch fixed point for 3 different Reynolds numbers, $\text{Re} =134.51,\ 134.52, \ 134.53$. The right panel show the curve of fixed points satisfying $J = R(J,\text{Re})$, obtained by continuation on the reduced model, that reproduces the saddle-node bifurcation of the full model. For numerical continuation, we use the MATLAB\textsuperscript{\tiny\textregistered} open source continuation core software \textsc{coco} \cite{coco}. In Fig. \ref{sfig3}, the mapping of the reduced dynamics in region (II) is plotted. For simplicity, we only plot one of the components, $R_{1}$, in the plane ($J,K$). 

Finally, we show a complete prediction result. First, we calculate the reduced trajectory for a random initial energy input rate $I_0$ at $\text{Re}=134.52$ in the region (I), then lift the trajectory to the full phase space via the parametrization. Figure \ref{sfig4} shows a comparison of the predicted and true velocity fields after $t=184$ time units. The streamwise ($u$) components of the velocity fields are shown in the $y=0$ plane. 
\subsection{Region (III)}
In region (III), we find that a reduced-order model based on a two-dimensional manifold containing all transitions cannot be constructed. This is because the unstable manifold of the lower branch spirals onto a limit cycle, and hence is not a differentiable manifold. To infer the minimal dimension of a higher-dimensional SSM containing the transition, we show in Table \ref{tab2} the leading eigenvalues of all coexisting ECSs at Re=146.

\begin{table}[h!]
\label{tab2}
\begin{center}
\begin{tabular}{ |c|c|c|c| } 
\hline
Base state & Lower branch & Upper branch & Limit cycle \\
 \hline
-0.0323110 & 0.0406734 & 0.0016963 $+\ i$ 0.0927744 & 0 \\ 
-0.0642890 & -0.0492400 $+\ i$ 0.1028837  & 0.0016963 $-\ i$ 0.0927744 & -0.0041905 \\ 
-0.0676000 & -0.0492400 $-\ i$ 0.1028837 & -0.0398964 $+\ i$ 0.0689366 & -0.0367628  $+\ i$ 0.0206478 \\
-0.1292439 & -0.0539350 & -0.0398964 $-\ i$ 0.0689366 & -0.0367628  $-\ i$ 0.0206478 \\
 \hline
\end{tabular}\label{tab:eigtableregionIII}
\end{center}
\caption{The leading four eigenvalues of the coexisting ECSs at Re=146. The eigenvalues are in descending order based on their real parts (decay rate). In the last column, the Floquet exponents of the limit cycle are reported. These are defined as $\lambda = \frac{1}{T}\log \Lambda$, where $\Lambda$ is the Floquet multiplier and $T$ is the period of the limit cycle.  }
\end{table}
Based on the spectrum of the limit cycle, the  slowest smooth SSM that captures the lower transition orbit would be four-dimensional.

Instead of a higher dimensional model, in region (III), we restrict our analysis to a neighborhood of the limit cycle, which allows us to represent the reduced dynamics via the polar normal form
\begin{equation}\label{eq:polarnormalform}
    \dot{\rho}= \sum_{n=0}^{ \text{Int } (M_d-1)/2 } c_n \rho^{2n+1}, \qquad \dot{\theta} = \sum_{n=0}^{ \text{Int } (M_d-1)/2 } d_n \rho^{2n},
\end{equation} 
with the variables $(\rho, \theta)$  linked to $(J,K)$ via a nonlinear change of coordinates. The model in Eq. (\ref{eq:polarnormalform}) is able to capture the transition from an unstable fixed point to an attracting limit cycle, and its form is reminiscent of the Hopf normal form \cite{guckenheimerNonlinearOscillationsDynamical1983} at first sight. Note, however, that in the classic Hopf normal form, the coefficient $c_0$  is a small bifurcation parameter, whereas in our setting, $c_0$  is not small and no closeness to a bifurcation is assumed.  Indeed, we do not perform the classic normal form procedure familiar from center-manifold reduction near non-hyperbolic fixed points. Instead, we use the data-driven extended normal form methodology outlined in \cite{cenedeseDatadrivenModelingPrediction2022} for hyperbolic fixed points, which we briefly review here for our particular setting.

After centering coordinates at the upper state, we first estimate the linear part $\mathbf{R}_1\in\mathbb{R}^{2\times 2}$ via regression for the dynamics in the coordinates $(J,K)$. In region (III), the eigenvalues of $\mathbf{R}_1 = \mathbf{P}\mathbf{D}\mathbf{P}^{-1}$, which are the entries of the diagonal matrix $\mathbf{D}$, are the complex conjugate pair with positive real part $c_0\pm id_0$, cf. Table 2. As this equilibrium is hyperbolic with nonresonant eigenvalues, the normal form on the SSM  would be purely linear  \cite{guckenheimerNonlinearOscillationsDynamical1983}. However, the domain of validity of such a linear normal form would be limited and could not capture the transition to a limit cycle. To this end, via the extended normal form procedure of \cite{cenedeseDatadrivenModelingPrediction2022}, we allow  for nonlinear terms (i.e., those with $n>0$ in Eq. (\ref{eq:polarnormalform}) that  would  arise in the reduced dynamics if the real part of the eigenvalues of the linearized spectrum on the SSM were zero at the fixed point. This enlarges the domain of validity of our data-driven reduced-order model that is now capable of capturing nonlinear behavior. For our particular case, we set $\mathbf{y} = \mathbf{P}^{-1}(J,K)$ and $\mathbf{z}=(z,\bar{z})$ with $z=\rho e^{i\theta}$, and we define the set $S(M_d):=\{ (l,m)\in\mathbb{N}^2:2\leq l+m\leq M_d,\,l\neq m+1\}$ to be used in the near-identity change of coordinates
\begin{equation}
    \mathbf{y} = z + \mathbf{h}(\mathbf{z}) = (h_1(z,\bar{z}),\bar{h}_1(z,\bar{z})), \qquad h_1(z,\bar{z}) = z + \sum_{(l,m)\in S(M_d) } h_{l,m}z^{l}\bar{z}^{m}.
\end{equation}
The inverse transformation, $ \mathbf{z} = \mathbf{h}^{-1}(\mathbf{y})$ features the same monomial structure. The dynamics takes the complex normal form of that in (\ref{eq:polarnormalform}), i.e.,
\begin{equation}
    \dot{\mathbf{z}} = \mathbf{n}(\mathbf{z}) = (n_1(z,\bar{z}),\bar{n}_1(z,\bar{z})), \qquad n_1(z,\bar{z}) = \sum_{n=0}^{ \text{Int } (M_d-1)/2 } (c_n+id_n)z^{n+1}\bar{z}^{n}.
\end{equation}
Note that the nonlinear monomials present in the normal form $n_1$ do not appear in the coordinate changes $h_1,h_1^{-1}$. We then find the nonlinear coefficients of the mappings $\mathbf{h}^{-1}$ and $\mathbf{n}$ via minimization of the conjugacy error
\begin{equation}
(\mathbf{h}^{-1*},\mathbf{n}^{*})= \mathrm{arg}\min_{\mathbf{h}^{-1},\mathbf{n}}\sum_{j=1}^{N_{dp}}\left\Vert \frac{d}{dt}\mathbf{h}^{-1}(\mathbf{y}_{j})-\mathbf{n}\left((\mathbf{h}^{-1}(\mathbf{y}_{j})\right)\right\Vert ^{2},
\label{eq:learningdynamics}
\end{equation}
for the $N_{dp}$ trajectory datapoints available. The nonlinear coefficients of $\mathbf{h}$ are estimated via regression afterwards, when the coordinates $\mathbf{z}$ are known. This extended normal form identification problem is solved automatically by the \texttt{SSMLearn} algorithm.
%